# Raman and Far Infrared Synchrotron Nanospectroscopy of Layered Crystalline Talc:

## Vibrational Properties, Interlayer Coupling and Symmetry Crossover

Raphael Longuinhos[1,7], Alisson R. Cadore[2,3,7], Hans A. Bechtel[4], Christiano J. S. de Matos[3,5], Raul O. Freitas[6], Jenaina Ribeiro-Soares[1+], Ingrid D. Barcelos[6+]

[1]Departamento de Física, Universidade Federal de Lavras (UFLA), Zip Code 37200-900, Lavras, Minas Gerais, Brazil.
[2]Present address: Brazilian Nanotechnology National Laboratory (LNNano), Brazilian Center for Research in Energy and Materials (CNPEM), Zip Code 13083-970, Campinas, Sao Paulo, Brazil.
[3]School of Engineering, Mackenzie Presbyterian University, Zip Code 01302-907, São Paulo, Brazil.
[4]Advanced Light Source (ALS), Lawrence Berkeley National Laboratory, Zip Code 94720, Berkeley, California, USA.
[5]MackGraphe, Mackenzie Presbyterian Institute, Zip Code 01302-907, São Paulo, Brazil.
[6]Brazilian Synchrotron Light Laboratory (LNLS), Brazilian Center for Research in Energy and Materials (CNPEM), Zip Code 13083-970, Campinas, Sao Paulo, Brazil.
[7]These authors contributed equally to this work.
+Corresponding-Author: jenaina.soares@ufla.br, ingrid.barcelos@lnls.br

## Abstract

Talc is an insulating layered material that is stable at ambient conditions and has high-quality basal cleavage, which is a major advantage for its use in van der Waals heterostructures. Here, we use near-field synchrotron infrared nanospectroscopy, Raman spectroscopy, and first-principles calculations to investigate the structural and vibrational properties of talc crystals, ranging from monolayer to bulk, in the 300-750 cm$^{-1}$ and <60 cm$^{-1}$ spectral windows. We observe a symmetry crossover from mono to bilayer talc samples, attributed to the stacking of adjacent layers. The in-plane lattice parameters and frequencies of intralayer modes of talc display weak dependence with the number of layers, consistent with a weak interlayer interaction. On the other hand, the low-frequency (<60 cm$^{-1}$) rigid-layer (interlayer) modes of talc are suitable to identify the number of layers in ultrathin talc samples, besides revealing strong in-plane and out-of-plane anisotropy in the interlayer force constants and related elastic stiffnesses of single crystals. The shear and breathing force constants of talc are found to be 66% and 28%, respectively, lower than those of graphite, making talc an excellent lubricant that can be easily exfoliated. Our results broaden the understanding of the structural and vibrational properties of talc at the nanoscale regime and serve as a guide for future ultrathin heterostructures applications.

## 1 – Introduction

Layered materials (LMs) possess exquisite electrical, optical, and mechanical properties that have been predicted theoretically[1–4] and measured experimentally by Raman spectroscopy[5–9], infrared spectroscopy[10–16], and other analytical techniques[17–20]. The potential use of LMs as building blocks for future ultrathin and flexible devices[21,22] has created interest from both scientific and economic perspectives to search for and characterize LMs that are easily obtained in nature[23–26]. Talc, also known as soapstone, is an abundant, naturally occurring magnesium hydrosilicate mineral from the phyllosilicate group, and is the softest known mineral[27]. It is an electrical insulator (bandgap of ~5 eV)[28,29] and allows for excellent basal cleavage, with layers held together by van der Waals forces[30,31], making it an excellent target for future low-cost optoelectronic applications. Figure 1a-c presents the optical images of the bulk crystal (mineral "block"); of an ultrathin staircase-like exfoliated sample of natural talc; and the schematic crystalline structure, respectively. Talc has the chemical formula $Mg_3Si_4O_{10}(OH)_2$ with a triclinic crystal lattice showing a 2:1 trioctahedral layered silicate structure (Fig. 1c), where one Mg-octahedral layer is covalently bound to two adjacent Si-tetrahedral layers (T-$O_c$-T structure)[30–32], which are henceforth referred to as a layer or a talc monolayer (1L-talc).

Monolayer (1L) and few-layer (FL) talc have recently revealed interesting magnetic[33], optical[34–36], electrical[28,29,37,38], and mechanical[32,39–43] properties as a single material and when combined with other LMs in van der Waals heterostructures (vdWHs). Although the structural properties of talc have been investigated previously[33,34,38–40,42,43], to the best of our knowledge, no work has fully described its Raman and far-infrared (Far-IR) active modes in the ultrathin form (Fig. 1d) as a function of layer thickness.

Here, we investigate the structural and vibrational properties of talc layers, from its ultrathin form to bulk scale using ab initio calculations, Raman spectroscopy, and synchrotron infrared nanospectroscopy (SINS). SINS enables a highly sensitive route to investigate samples with reduced lateral size (<100 nm)[10,11] due to its tip-driven spatial resolution (<25 nm) and broad spectral range coverage (330 cm$^{-1}$ to 4000 cm$^{-1}$[44,45], in contrast to conventional Fourier-transform IR (FTIR) spectroscopy, which has a diffraction-limited spot size that is too large to investigate typical individual talc flakes[46]. We use symmetry analysis to rationalize the structural dependence with the number of layers (N-layers) of talc lattice modes and geometry, showing that a symmetry crossover occurs from the 1L to bilayer (2L) talc. Further insight on atomic displacement pattern of the Far-IR modes of talc is obtained by first-principles calculations, and we show that the low-frequency (<60 cm$^{-1}$) Raman modes can be used to assess the number of layers in talc flakes. Finally, we study the interlayer force constants in layered talc and reveal that single crystals display anisotropic in-plane and out-of-plane forces constants, with values indicating high lubricity and corroborating the ease of exfoliation. Therefore, our results serve as a guide for future studies, allowing for the identification of the exact structure of ultrathin layered talc, and facilitating further structural customizations, as well as its use in vdWHs applications.

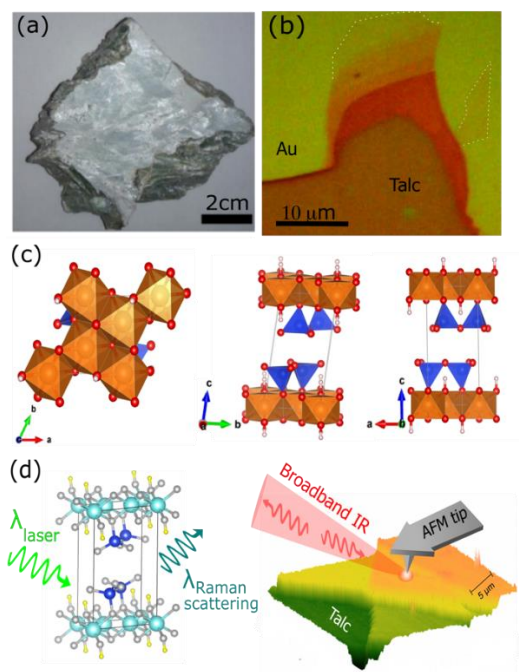

**Figure 1:** Optical characterization scheme and talc layered structure. (a) Mineral talc obtained from a talc/soapstone mine in Ouro Preto (Brazil). (b) False-color optical microscope image of an exfoliated staircase-like talc flake atop a gold substrate. (c) Atomistic model structure of bulk talc, where the basal plane is defined by the lattice vectors **a** and **b**, in the *xy*-plane, and the stacking direction is defined by the lattice vector **c**. (d) Overview of the experimental techniques used to probe the vibrational properties of talc nanocrystals: Raman spectroscopy (left) and Far-IR SINS (right).

## 4 – Methods

### 4.1 – Experimental section

The natural bulk-talc sample used in this work (Fig. 1a) was obtained from a mine located in Ouro Preto, Brazil. Standard mechanical exfoliation was used to produce ultrathin and staircase-like talc flakes on top of a gold film (100nm), deposited on prime silicon substrate.

Raman spectroscopy was carried out in a confocal Raman microscope WITec Alpha 300R with spectral resolution of ~0.5 cm$^{-1}$ from a freshly exfoliated talc crystal on a glass substrate in the backscattering configuration. We used 488 nm excitation energy with a 100x objective, which has a numerical aperture (NA) of 0.9. The laser power was kept constant at ~20 mW during all measurements. The laser was aligned on a prime silicon wafer, using the Si peak at ~521 cm$^{-1}$ as a reference[47]. Each spectrum was then collected from 50 cm$^{-1}$ to 4000 cm$^{-1}$ with three accumulations of 60 sec.

SINS experiments were performed at the Advanced Light Source (ALS) at the Lawrence Berkeley National Lab. The SINS technique combines a high-brightness, broadband synchrotron light source (frequency window is 330-4000 cm$^{-1}$), with a FTIR spectrometer and an AFM probe[10,11]. The beamline optical setup consists of an asymmetric Michelson interferometer mounted into a commercial s-SNOM microscope (NeaSnom, NeaspecGmBH), which can be described as an AFM with suitable optical access to excite and collect the near-field scattered light. In the interferometer, the incident synchrotron IR beam is split in two components by a KRS-5 beamsplitter defining the two interferometer arms, consisting of a metallic AFM tip (tip arm) and an IR high-reflectivity mirror mounted onto a translation stage (scanning arm). The IR beam component of the tip arm is focused by a parabolic mirror on the tip-sample region. In the experiment, the AFM operates in semi-contact (tapping) mode, wherein the tip is electronically driven to oscillate in its fundamental mechanical frequency $\Omega$ (~250 kHz) near the sample surface. The incident light interacts with the metallic coated tip and sample, creating an effective local polarization. The back-scattered light stemming from this tip-sample interaction, composed of far- and near-field contributions, is collected by a high-speed IR detector and a lock-in amplifier having $\Omega$ as the reference frequency. The clean optical near-field ($S_n$), originated from a sample area of radius comparable to the tip radius (around 25 nm in our case), is given by the back-scattered light components modulated in high harmonics (n) of $\Omega$, with $n \geq 2$. Displacing the mirror in the scanning arm in this asymmetric scheme creates complex interferograms that can be Fourier Transformed to yield amplitude $|S_n(\omega)|$ and phase $\Phi_n(\omega)$ spectra. All SINS spectra here are given by $n = 2$, i.e., $S_2(\omega)$. For the Far-IR measurements at ALS Beamline 2.4, a customized Ge:Cu photoconductor, which provides broadband spectral detection down to 320 cm$^{-1}$, and a KRS-5 beamsplitter were employed. Therein, we obtain 20 SINS spectra per point in our measurements, which are averaged and normalized to a reference spectrum obtained from a pure gold surface.

### 4.2 – Calculations

We applied first-principles calculations within density functional theory and density functional perturbation theory[48–50]. FL and 1L-talc were theoretically considered by using a slab model and appropriate boundary conditions for 2D crystals[51,52], with vacuum equal to approximately 16 angstroms. All calculations were performed using Quantum Espresso distribution[53,54]. The structures were optimized until the forces on the atoms and stress on the lattice were lower than 2.57 meV/angstrom and 50 MPa, respectively, with a Brillouin zone sampling by using the Monkhorst-Pack scheme[55], with 4x4x2 and 4x4x1 Gamma center k-point grids for bulk- and FL-talc, respectively. In the linear response calculations[56], to obtain the Raman and IR spectra simulations, the Brillouin zone sampling was calculated by using 6x6x4 and 6x6x1 for bulk and FL-talc, respectively. The valence electron-nucleus interactions were described by using Troullier-Martin norm-conversing pseudopotentials[57], available at Quantum Espresso repositories, setting the kinetic energy cutoff in the wave-function (charge-density) expansion to 46 Eh (184 Eh). We considered the Perdew-Wang's parametrization of the local density approximation for the exchange-correlation (XC) functional[58]. In addition, for bulk, we also considered

PBEsol[59] and RVV10[60] XC functionals, where the former is an improved parametrization for solids of the Perdew Burker Ernzerhof PBE XC functional[61] and the latter is an improved implementation of the VV10 XC functional[62], which describes the van der Waals interactions. For estimation of the graphene stiffness coefficients here reported, we built the related stress *vs* strain curves by using the LDA XC functional and Troullier-Martin norm-conserving pseudopotentials, with kinetic energy cutoff in the wave-function (charge-density) expansion equal to 30 Eh (120Eh). The Brillouin zone sampling was performed by using the cold-smearing method, with electronic temperature equal to 15mEh, and 8x8x6 Gamma center within the Monkhorst-Pack method.

## 2 – Results and Discussion

### 2.1 – Crystal structure and lattice vibrations

Figure 1c displays the atomistic model for bulk talc, with 21 atoms within the primitive cell, given by $a = b = 5.211$ Å, $c = 9.187$ Å, $\alpha = 79.94°$, $\beta = 84.17°$, and $\gamma = 60.16°$, obtained from density-functional theory (DFT) calculations by using the local-density approximation (LDA) for the exchange-correlation functional (see Methods and Supporting Information for more details). The corresponding conventional cell has lattice parameters equal to $a = 5.211$ Å, $b = 9.039$ Å, $c = 9.187$ Å, $\alpha = 89.05°$, $\beta = 100.06°$, and $\gamma = 89.84°$, in excellent agreement to the experimental data[34], given by $a = 5.291$ Å, $b = 9.172$ Å, $c = 9.455$ Å, $\alpha = 90.58°$, $\beta = 98.77°$, $\gamma = 89.99°$. In addition, we found as a result of our DFT-LDA calculations that the talc layer has a mass per unit area equal to $2.67 \times 10^{-26}$ kg/Å$^2$, a thickness equal to 6.452 Å, and is apart by 2.593 Å from adjacent layers.

The symmetry analysis of the simulated talc structures shows that the 1L-talc is in the $C_{2h}^3$ space group, with inversion center, mirror, and rotation, whereas 2L-talc to bulk exhibit only an inversion center and are classified as $C_i^1$ structures. Thus, a symmetry crossover occurs in talc with the addition of one layer from 1L to 2L thickness, attributed to the stacking of adjacent layers. Table 1 shows the calculated in-plane lattice parameter, which has only a minor dependence with the number of layers (N-layers), which is consistent with the weak interlayer interactions. In the Supporting Information we compare the results for the bulk talc structure obtained by using other approximations for the exchange-correlation functional within DFT calculations and discuss the trends.

Table 1: Lattice parameter and space group of talc for the different number of layers (NL):

|  | a (Å) | space group |
|---|---|---|
| 1L | 5.224 | 12 ( $C_{2h}^3$ ) |
| 2L | 5.217 | 2 ( $C_i^1$ ) |
| 3L | 5.215 | 2 ( $C_i^1$ ) |
| 4L | 5.214 | 2 ( $C_i^1$ ) |
| bulk | 5.211 | 2 ( $C_i^1$ ) |

The symmetry dependence on the N-Layer has implications on the irreducible representations for the lattice vibrations in talc. The lattice modes of 1L-talc are composed by

$$\Gamma = 14A_u \oplus 19B_u \oplus 16A_g \oplus 14B_g \quad \text{Eq. 1.}$$

The $A_u$ and $B_u$ vibrational modes are IR active, while the $A_g$ and $B_g$ modes are Raman active. As N-layer increases (and the number of atoms in the unit cell), the expected number of modes also enhances. For a N-layer system, in the case of even N, the lattice modes are composed as

$$\Gamma = \frac{63N}{2}\left(A_u \oplus A_g\right) \quad \text{Eq. 2,}$$

while for odd N (excluding the N=1 previous case), it is given by

$$\Gamma = \frac{3}{2}\left((21N+1)A_u \oplus (21N-1)A_g\right) \quad \text{Eq. 3.}$$

In the case of bulk talc, the lattice modes composition is

$$\Gamma = 33A_u \oplus 30A_g \quad \text{Eq. 4,}$$

and, similarly to the odd-N and even-N cases, the $A_u$ vibrational modes are IR active, while the $A_g$ modes are Raman active. All structures are centrosymmetric, so Raman and IR-active modes are mutually exclusive, as can be observed in the experimental and theoretical results in the next sections. These results unveil the expected number of vibrational modes in talc nanostructures and their optical activity.

## 2.2 – Far-IR investigation

Figure 2 displays the nanoscale IR vibrational analysis of the talc bulk crystal in the 300-750 cm$^{-1}$ spectral window. Figure 2a shows the experimental near-field amplitude $S_2(\omega)$ and phase $\varphi_2(\omega)$ IR spectra (see Methods for more details), with experimentally determined peak positions of the phase spectra indicated as blue stars. Many of these peaks (see Table 2) are new features that have not been detected in prior conventional IR studies of powder bulk talc[31,63]. Figure 2b shows the simulated spectrum for powder bulk talc, obtained by using density-functional perturbation theory (DFPT) calculations, where the solid line is the result of the Gaussian convolution of the relative IR intensity of the IR-active modes and the red circles indicate the wavenumber of the calculated IR-active modes of talc single crystal.

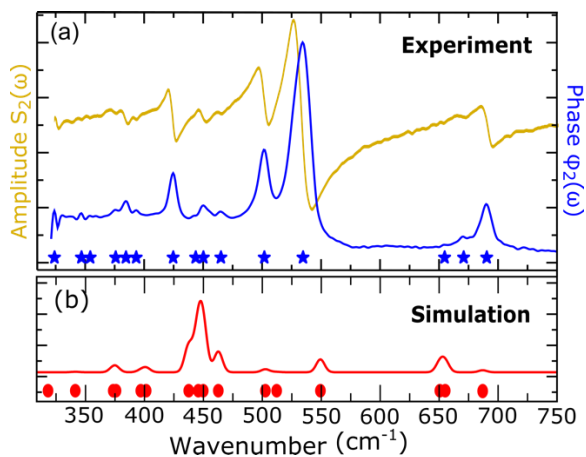

**Figure 2:** Far-IR spectra of bulk talc. (a) Near-field amplitude (yellow line) and phase spectra (blue line) with marked stars assigning the mode positions. (b) Simulated spectrum (blue line) for powder bulk talc, where the circles indicate the peak positions, and the solid line is the Gaussian convolution of the relative intensities.

We notice excellent agreement between DFPT simulation and conventional IR spectra of powder talc measured in previous works[31,63] (see Figure S2 in the Supplementary Information for the superposition of both simulated and experimental powder bulk spectra). Although the calculated frequencies agree well with those observed in the SINS measurements, the relative intensities of the simulated and experimental near-field spectra do not agree, and in fact, appear to be anticorrelated in the sense that the weaker features in the simulation have stronger intensities in the SINS spectra.

*Table 2: Simulated and experimental Infrared modes of bulk talc in the 300–750 cm$^{-1}$ range (see Supporting Information for more vibrational modes). All mode positions are in cm$^{-1}$. The IR resonances in bold were not detected in previous conventional-IR studies of powder bulk talc[31,63]. In some cases, there is more than one simulated vibration mode near in wavenumber from an experimental IR-resonance peak.*

| IR | Simulated IR | IR | Simulated IR |
|---|---|---|---|
| **323.9** | 318.2 | 450.2 | 449.9 |
| 346.8 | 341.3 | 464.8 | 462.6 |
| **354.1** | | 501.7 | 502.6; 512.2 |
| **375.8** | 373.5; 375.9 | 534.5 | 549.3 |
| 384.3 | 397.0 | **654.3** | 650.6 |
| 393.1 | 401.5 | 670.5 | 655.2 |
| 424.5 | 437.6 | 690.1 | 687.0 |
| 442.0 | 445.7 | | |

To further explore this behavior, we calculated the atomic displacement patterns of representative IR active vibration modes of talc as shown in Figure 3 and correlate these motions to the intensity patterns in the SINS phase spectrum. Figure 3a-d illustrates the Mg-O, O-H and Si-O bending modes (300-380 cm$^{-1}$); the Mg-O and O-H bending and stretching modes (400-460 cm$^{-1}$); the Mg-O, O-H and Si-O bending and stretching modes (500-550 cm$^{-1}$); and the O-H bending and stretching mode (~655 cm$^{-1}$) and Mg-O and Si-O stretching mode (~687 cm$^{-1}$), respectively. We find that the most prominent bulk talc mode in the SINS phase spectrum at 534.1 cm$^{-1}$, corresponds to essentially out-of-plane atomic movements (549.27 cm$^{-1}$ in Figure 3c).

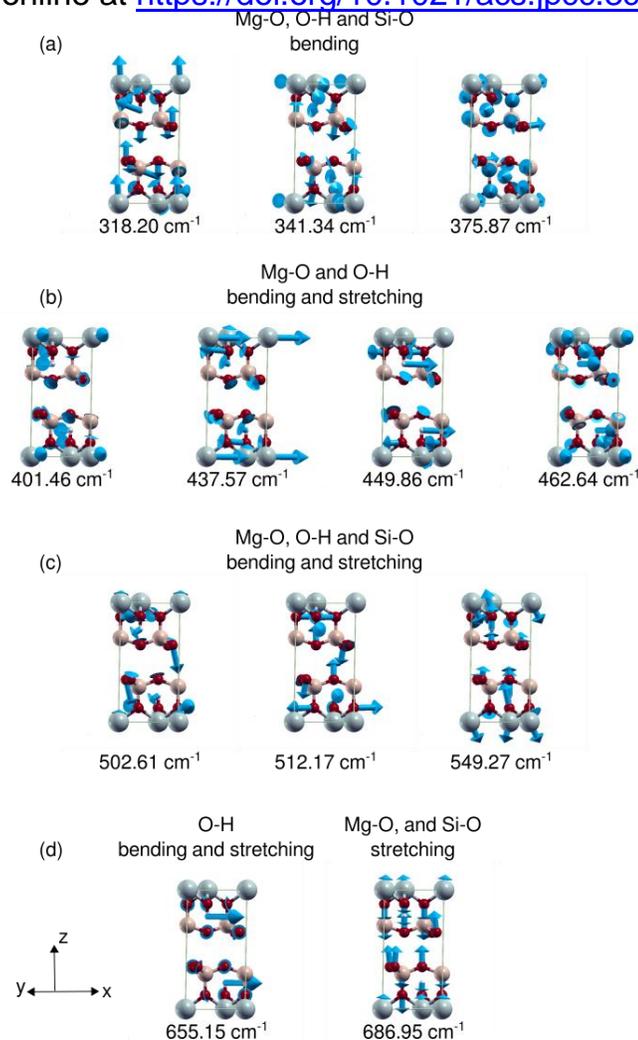

**Figure 3:** IR vibrational identity for talc. (a-d) Simulated atomic displacement patterns of representative Far-IR vibrations in bulk talc.

Conversely, the weaker features in the SINS phase spectrum correspond to mostly in-plane atomic movements. This strong signal enhancement for vibrations with out-of-plane components and signal diminution for vibrations with in-plane components arise from the experimental geometry, in which the basal plane is oriented perpendicular to the strongly polarized near-field electromagnetic mode in the atomic force microscope (AFM) tip, which preferentially excites dipole moment oscillations parallel to the tip axis (perpendicular to the basal plane), as observed in other-oriented samples [31]. The sensitivity of SINS to out-of-plane modes contrasts with conventional IR transmission measurements, which are primarily sensitive to in-plane modes because radiation propagating perpendicular to the talc basal plane cannot couple effectively to the out-of-plane vibrational modes. This orientation-related sensitivity difference between the two techniques, thus, accounts for the apparent anti-correlated intensity patterns of the SINS measurements and the simulated powder spectra. Indeed, the most prominent bulk talc mode in SINS is practically nonexistent in conventional absorption spectra of oriented sample deposited on KBr windows[31,64].

Figure 4 shows the SINS amplitude and phase spectra as a function of talc layers. The high-resolution AFM topography image (Fig. 4a), automatically obtained with the SINS setup[10,11], is used to determine the thickness of the exfoliated talc flakes. Simultaneously, the broadband near-field image, which accounts for broadband local reflectivity, is acquired with sharp optical contrast for all N-layers (Fig. 4b), allowing for precise identification of candidate positions for the SINS point spectra, labelled P1, P2, …, P8. Our calculations indicate that the 1L-talc thickness is approximately equal to 0.9 nm. Consequently, P1 corresponds to a 2L-talc (2 nm), P2 to a trilayer (3 nm), while other positions rise all way up to bulk (P8). The point-spectra of talc sample at position P8 in Figure 4c-d presents several clear resonances in the spectral range studied and these modes are shown in Table 2. The predominance of out-of-plane modes, attributed to the tip-enhanced nature of the near-field technique, is preserved for the nanoscale thicknesses down to the 2L-talc sample. The reduction in layer number (sample thickness) causes a decrease in the overall intensity, and the modes at 384.3 cm$^{-1}$, 450.2 cm$^{-1}$, and 464.8 cm$^{-1}$ get closer to noise level. Nevertheless, we still observe four clear resonances in the point spectrum acquired at P1 (2L-talc) at 425.2 cm$^{-1}$, 502.2 cm$^{-1}$, 537.8 cm$^{-1}$ and 690.6 cm$^{-1}$, which can be used as a vibrational trend signature of talc, from bulk to the 2L regime.

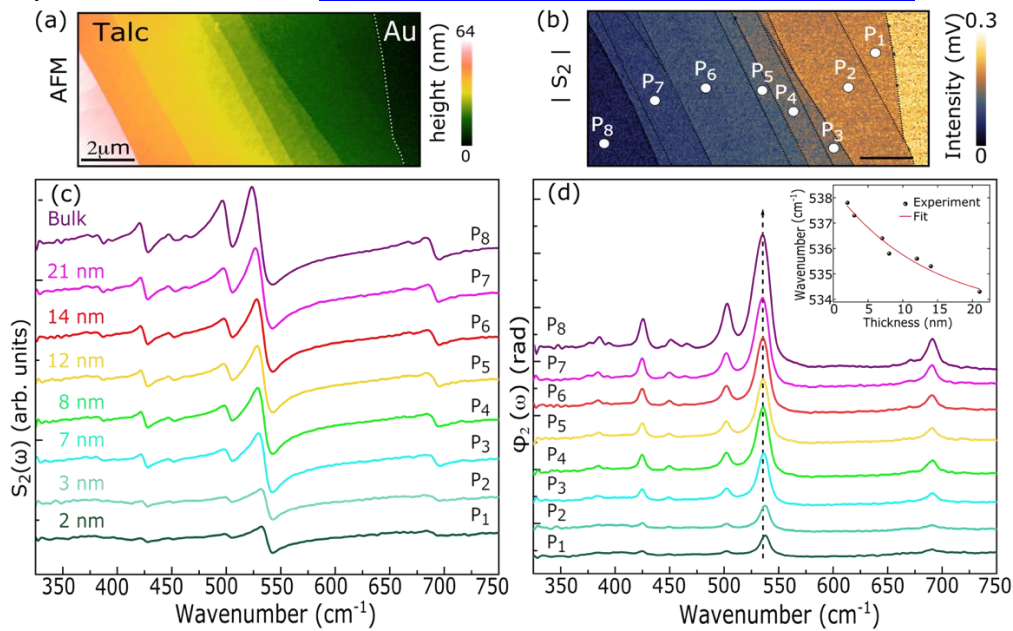

**Figure 4:** Infrared fingerprints of talc layers. (a) AFM topography image of a staircase-like FL-talc atop a gold substrate. The dashed line marks the edge of the talc flake and gold. (b) Broadband near-field reflectivity image of the same region revealing scattering intensity modulation according to the different talc N-layers. Black dashed lines guide the eye to the talc layers' boundaries. SINS point spectra: near-field (c) amplitude $S_2(\omega)$ and (d) phase $\phi_2(\omega)$ of talc crystal with different thicknesses. The overall spectral range focuses on the available near-field frequency window (330-750 cm$^{-1}$) of interest. Inset: Empirical analysis of the mode at ~534 cm$^{-1}$, where spheres indicate the experimental data taken from (d), while the continuous-red line indicates the fit.

The wavenumber dependence of these resonances with the number of layers is found to be weak. However, the prominent resonance at around 537.8 cm$^{-1}$ in the 2L-talc, which shifts towards 534.4 cm$^{-1}$ in bulk talc, can be used to estimate the layer number in ultrathin talc samples as shown in the inset of Figure 4d. We have then defined an empirical relation for a ready-to-use layer number determination strategy: $w = w_0 + A * exp\ (-B * t)$, where $\omega_0$ = 533.33 cm$^{-1}$ is the bulk wavenumber value (SINS equal to 534.4cm$^{-1}$), $A$ = 5.034 cm$^{-1}$ and $B$ = 0.073 nm$^{-1}$, and '$t$' the material thickness in nanometers (we found a correlation coefficient equal to 0.964 for this fit). The wavenumber of this resonance decreases as N-layer increases, contrary to what is predicted from the simulations (see Supporting Information). This experiment-theory contradiction in mode shift trend has been previously noticed for other materials in SINS-DFPT combined studies[65] and indicate the role of physical effects beyond those described by DFPT[66–68].

## 2.3 – High-frequency Raman investigation

Figure 5a presents the representative Raman spectrum of bulk talc, collected with a 488 nm excitation laser. The wavenumber of the Raman modes detected here are shown in Table 3. The Raman peaks at 194.5 cm$^{-1}$, 362.1 cm$^{-1}$, 676.2 cm$^{-1}$ and 1050.5 cm$^{-1}$ are ascribed to the fundamental vibrations of silicates in talc, and the peak at 3678.8 cm$^{-1}$ is the characteristic OH mode of talc hydroxyl groups bound to the talc octahedral layer[69]. The additional peak at 3663.4 cm$^{-1}$ may indicate a slight octahedral distortion[69], which leads to the splitting of the MgO-OH mode, or iron contamination in the talc octahedral layer[70,71]. However, the latter is less likely to be the cause, as the measured Raman shift of the talc fundamental peaks of our sample agree with that of high-quality standard talc samples, without any indication of iron impurities[69] or the three characteristic Fe-OH modes near 3678 cm$^{-1}$ in iron-rich talc[71]. Figure 5b shows the DFPT-simulated Raman spectrum of bulk talc, showing excellent agreement between theory and the experiment (see Table 3 and the Supporting Information, where we compare the results for the bulk talc Raman active modes obtained by using PBEsol and rVV10 functionals).

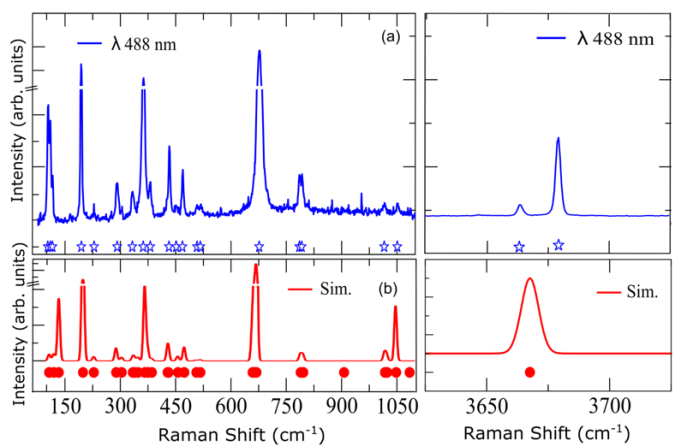

**Figure 5:** Raman spectroscopy characterization. (a) Raman spectrum of a bulk talc flake collected using a 488 nm excitation laser. (b) Simulated spectrum for powder bulk talc, where the circles indicate the peak position, and the solid line is the Gaussian convolution of the relative intensities.

*Table 3: Simulated and experimental Raman modes of bulk talc. All mode positions are in cm$^{-1}$. In some cases, there is more than one simulated vibration mode near in wavenumber from an experimental Raman scattering peak. The Raman peak marked by * is from Ref.[72].*

| Experiments | Simulation | Experiments | Simulation |
|---|---|---|---|
| 103.8 | 106.9 | 468.4 | 473.4 |
| 110.8 | | 508.0 | 506.2 |
| 116.1 | 119.4 | 517.1 | 517.0 |
| 132* | 133.1 | 676.2 | 658.7; 661.9; 668.8 |
| 194.5 | 198.7 | 785.2 | 788.9 |
| 229.0 | 228.0 | 791.8 | 795.9 |
| 291.5 | 288.6 | | 906.6 |
| 304.8 | 303.6 | 1014.9 | |
| 332.6 | 333.9; 338.0 | 1018.3 | 1017.5 |
| 345.8 | 348.1 | 1020.7 | 1023.4 |
| 362.1 | 366.3 | 1050.5 | 1047.7 |
| 370.3 | 374.4 | 1085.2 | 1084.3 |
| 381.1 | 385.4 | 3663.4 | 3667.6 |
| 431.9 | 428.2; 430.5 | 3678.8 | |
| 451.6 | 456.1 | | |

The spectral window from 100 cm$^{-1}$ to 1100 cm$^{-1}$ in Figure 5a, such as fundamental modes in silicates, is dominated by rigid translation and rotation of the TOT modules (80-250 cm$^{-1}$), lattice modes involving Mg–O bonds and tetrahedral bending (250-650 cm$^{-1}$), strong bending and stretching modes of Si–O–Si bridges, around 670 cm$^{-1}$ and above 1000 cm$^{-1}$, respectively[69,73–75]. The peaks in the 3300-3700 cm$^{-1}$ spectral range in Figure 5a-b, for example, H$_2$O/OH vibrations, correspond to stretching modes of OH linked to octahedral sites[69,73–75]. Figure 6 displays the calculated atomic displacement patterns of representative Raman active modes of talc in these spectral windows.

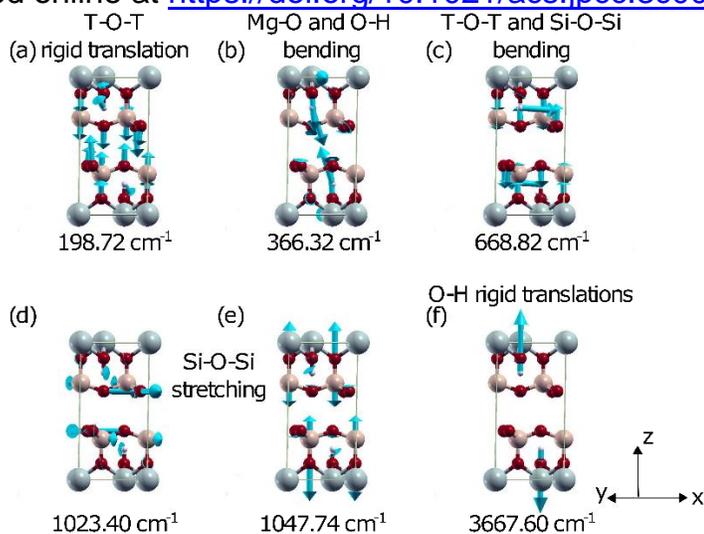

**Figure 6:** Raman vibrational identity for talc. (a-f) Atomic displacement patterns of representative Raman-active modes in bulk-talc.

No clear Raman response is observed within the experimental signal-to-noise ratio (SNR) for FL-talc flakes (only the three most intense modes are observed[43] around 194.8 cm$^{-1}$, 676.2 cm$^{-1}$, and 3678.8 cm$^{-1}$ for exfoliated bulk-talc flakes). Thus, we considered the simulation results to investigate the Raman wavenumber dependence on the talc layer thickness. The simulated Raman spectra for powder-layered talc is shown in Figure 7, where the 1L-talc spectrum is at the bottom, and the 2L, 3L, 4L and bulk talc (upper curves, respectively) spectra are displaced vertically. The bulk and FL-talc spectra present a similar pattern, showing that we can use the talc fundamental modes as fingerprints of the transition from 1L to FL- talc (see more details in the Supporting Information). It is worth noticing that as the number of layers increase, the number of vibrational modes also increases (see the discussion in section Crystal structure and lattice vibrations), and the total Raman (and SINS, see Figure 4 for measurements and SI for IR simulations) spectrum peak is formed by the convolution of these emerging modes at the vicinity of the modes observed in the bulk counterpart[76].

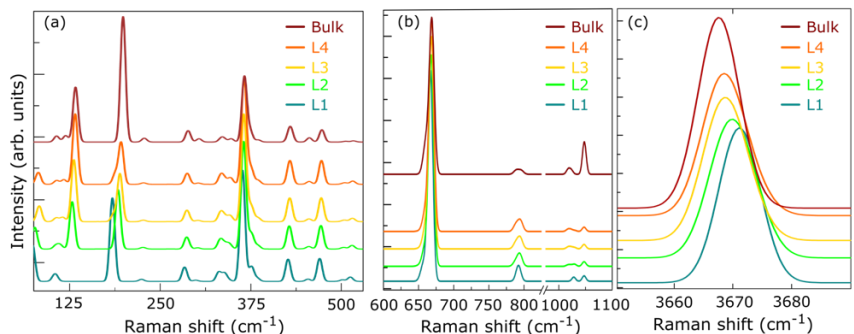

**Figure 7:** Trends of the Raman shift for powder-layered talc, from 1L to bulk. Solid lines are the Gaussian convolution of the relative intensities of the Raman-active modes. (a), (b) e (c) depicts the Raman spectra in the 75-550 $cm^{-1}$, 600-1100 $cm^{-1}$, and 3650-3690 $cm^{-1}$ spectral window, respectively.

## 2.4 – Low-frequency modes

The low-frequency (<60 $cm^{-1}$), rigid-layer modes in LMs, where adjacent layers move rigidly with respect to one another and without intralayer optical vibrations (>100 $cm^{-1}$), can be used to probe the sample thickness and interlayer interactions in LMs[77–80]. Figure 8 shows the simulated low-frequency interlayer shear and breathing modes for 2L, 3L, 4L, and bulk talc, following the branch where adjacent layers present an out-of-phase displacement. In Figure 8a-b, the shear displacement is along the armchair and zigzag direction of the silicon-like hexagons in the tetrahedral silicate layer, respectively. In Figure 8c, the breathing displacement is perpendicular to the layer plane. The wavenumber difference between the 2L- and 3L-talc interlayer shear modes along the armchair-like direction is 2.3 $cm^{-1}$. In the case of the interlayer shear mode along the zigzag-like direction, this difference is 7.3 $cm^{-1}$. For the interlayer breathing mode, this difference is also approximately 7.3 $cm^{-1}$.

In addition, the wavenumber difference between the 2L and bulk talc interlayer shear modes along the armchair direction is 4.58 $cm^{-1}$. In the case of the interlayer shear mode along the zigzag direction, this difference is 15.42 $cm^{-1}$. For the interlayer breathing modes, this difference is 17.90 $cm^{-1}$. Thus, the wavenumber dependence between the layers in this spectral range are much larger than the ones for >100 $cm^{-1}$, consequently, this could be used to assign talc N-layers. We notice that measuring the low-frequency $A_u$ modes (IR-active) is beyond the current nano-IR spectroscopy technology[10,11] and in the case of bulk, these modes are not zone-centre modes. However, measuring the low-frequency $A_g$ modes (Raman-active) is currently possible[79,81,82].

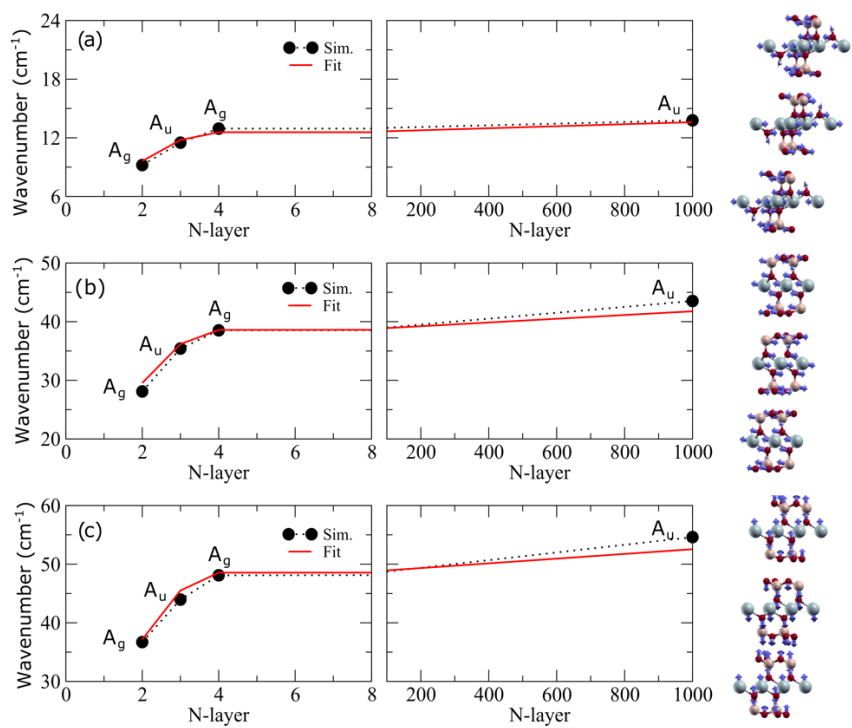

**Figure 8:** Low-frequency Raman modes in layered talc. (a) Shear mode along the armchair direction. (b) Shear mode along zigzag direction. (c) Breathing mode along stacking direction. Closed circles are first-principles calculation results and solid red lines are the fittings of the linear-chain modes to these points. The atomic displacement pattern of the modes is shown at the right of the graphics.

In the previous sections, we found the wavenumber dependence of the optical vibrations (>100 cm$^{-1}$) with Raman and IR activity to N-layers to be weak. As an example, good candidates for the assessment of the optical vibrations regarding the sample thickness are: i) the IR-active mode with both the strongest relative resonance and the strongest wavenumber dependence with the thickness in the Far-IR window detected at 537.8 cm$^{-1}$ in the 2L-talc (that shifts up to 3.4 cm$^{-1}$ towards lower wavenumber from 2L to bulk, as shown in Figure 4d); ii) the "fundamental" Raman-active modes, with scattering peaks at calculated average wavenumbers (the average consider the wavenumber and relative intensity of modes that contribute to the peak) equal to 191.8 cm$^{-1}$, 365.2 cm$^{-1}$ and 668.0 cm$^{-1}$ in 2L-talc (equal to 194.5 cm$^{-1}$, 365.5 cm$^{-1}$ and 668.2 cm$^{-1}$ in 3L-talc, Δω of 2.7 cm$^{-1}$, 0.3 cm$^{-1}$ and 0.2 cm$^{-1}$, respectively, and equal to 198.7 cm$^{-1}$, 366.3 cm$^{-1}$, and 668.8 cm$^{-1}$ in bulk talc, $\Delta\omega$ of 6.9cm$^{-1}$, 1.1cm$^{-1}$ and 0.87 cm$^{-1}$, respectively) as shown in Figure 7a-b. Thus, we propose the use of the wavenumber dependence of the low-frequency modes to on N-layers as a major signature for the sample thickness.

The low-frequency rigid-layer modes in LMs can be rationalized by the application of a linear-chain model[8,77–81,83,84]. By fitting the linear-chain model to our calculation results (red line in Figure 8), we can obtain the interlayer force constants of layered talc, from bulk to the 2L regime. We find that the interlayer force constant for shear movement ($K_{LSM}$) along the armchair- and zigzag-like directions are 0.44 x10$^{19}$ N/m$^3$ and 4.14 x10$^{19}$ N/m$^3$, respectively. The in-plane anisotropy for the shear movement is about 840%, showing that talc should slide preferentially in the armchair direction in comparison to the zigzag. The $K_{LSM}$ along the armchair direction in talc is about 66% smaller than that in graphene and GaSe[80] and about 91% smaller than that in jacutingaite and tilkerodeite minerals[84]. The $K_{LSM}$ along the zigzag direction in talc is about 218% greater than that in graphene[79] and GaSe[80] and about 20% smaller than that in jacutingaite and tilkerodeite minerals[84]. In the case of the interlayer breathing force constant ($K_{LMB}$), we found 6.55 x10$^{19}$ N/m$^3$, which is about 28% smaller than that in graphene[79], and about 45% and 52% greater than that in b-GaSe and e-GaSe[80], respectively, and about 65% smaller than that in jacutingaite and tilkerodeite minerals[84]. These results agree with the fact that talc can be easily exfoliated and is an excellent lubricant[32,39,43].

Finally, we extract the talc interlayer stiffness coefficients against interlayer shear movement along armchair and zigzag-like directions, and against strain perpendicular to the layers, given by 4.0 GPa, 37.5 GPa and 59.3 GPa, respectively. We estimate for graphite a $C_{44}$ equal to 5.8 GPa and a $C_{33}$ equal to 36.5 GPa, which is consistent with reported values[85]. The interlayer shear stiffness along the armchair in talc is comparable to that of graphite, while its interlayer shear stiffness along the zigzag direction is about 546% greater than that of graphite. In addition, talc is 63% stiffer than graphite against strain perpendicular to its basal plane. Although the $K_{LBM}$ of talc is smaller than that of graphite, a talc layer is 177% thicker than a graphite layer, showing that both thickness and $K_{LBM}$ need to be accounted for to understand the macroscopic elasticity of materials in the linear-chain model.

## 3 – Conclusion

We investigated layered talc's structural and vibrational properties, from bulk to monolayer regime. The in-plane lattice parameters of talc display negligible dependence on the number of layers, characteristic of weak interlayer interactions. We found the stacking of layers in talc to reduce its symmetries, where a symmetry cross-over occurs from mono ($C_{2h}^3$ space group) to bilayer-talc ($C_i^1$), leading to changes in the symmetries of the lattice modes. We determined the Far-IR response of ultrathin talc, by SINS revealing new features in comparison to the conventional approach. We then presented Far-IR talc modes general atomic displacement patterns, obtained by first-principles calculations. In addition, Raman spectroscopy combined with first-principles calculations complemented the knowledge of the lattice vibrations of talc in the 300-750 cm$^{-1}$ window, from monolayer to bulk. We also showed that the low-frequency (<60 cm$^{-1}$) rigid-layer modes were more sensitive to the variation of the number of layers than the optical layer modes (>100 cm$^{-1}$) and that they can be used to infer the number of layers in talc flakes. Finally, we studied the interlayer force constants in layered talc and revealed that single crystals displayed anisotropic in-plane and out-of-plane forces constants, with values indicating high lubricity and that it can be easily exfoliated. Therefore, our results broaden the understanding of the structural and vibrational properties of talc at the nanoscale, facilitating its use in ultrathin vdWHs applications.

## 4 – Methods

### 4.1 – Experimental section

The natural bulk-talc sample used in this work (Fig. 1a) was obtained from a mine located in Ouro Preto, Brazil. Standard mechanical exfoliation was used to produce ultrathin and staircase-like talc flakes on top of a gold film (100nm), deposited on prime silicon substrate.

Raman spectroscopy was carried out in a confocal Raman microscope WITec Alpha 300R with spectral resolution of ~0.5 cm⁻¹ from a freshly exfoliated talc crystal on a glass substrate in the backscattering configuration. We used 488 nm excitation energy with a 100x objective, which has a numerical aperture (NA) of 0.9. The laser power was kept constant at ~20 mW during all measurements. The laser was aligned on a prime silicon wafer, using the Si peak at ~521 cm⁻¹ as a reference[47]. Each spectrum was then collected from 50 cm⁻¹ to 4000 cm⁻¹ with three accumulations of 60 sec.

SINS experiments were performed at the Advanced Light Source (ALS) at the Lawrence Berkeley National Lab. The SINS technique combines a high-brightness, broadband synchrotron light source (frequency window is 330-4000 cm⁻¹), with a FTIR spectrometer and an AFM probe[10,11]. The beamline optical setup consists of an asymmetric Michelson interferometer mounted into a commercial s-SNOM microscope (NeaSnom, NeaspecGmbH), which can be described as an AFM with suitable optical access to excite and collect the near-field scattered light. In the interferometer, the incident synchrotron IR beam is split in two components by a KRS-5 beamsplitter defining the two interferometer arms, consisting of a metallic AFM tip (tip arm) and an IR high-reflectivity mirror mounted onto a translation stage (scanning arm). The IR beam component of the tip arm is focused by a parabolic mirror on the tip-sample region. In the experiment, the AFM operates in semi-contact (tapping) mode, wherein the tip is electronically driven to oscillate in its fundamental mechanical frequency $\Omega$ (~250 kHz) near the sample surface. The incident light interacts with the metallic coated tip and sample, creating an effective local polarization. The back-scattered light stemming from this tip-sample interaction, composed of far- and near-field contributions, is collected by a high-speed IR detector and a lock-in amplifier having $\Omega$ as the reference frequency. The clean optical near-field ($S_n$), originated from a sample area of radius comparable to the tip radius (around 25 nm in our case), is given by the back-scattered light components modulated in high harmonics (n) of $\Omega$, with n ≥ 2. Displacing the mirror in the scanning arm in this asymmetric scheme creates complex interferograms that can be Fourier Transformed to yield amplitude $|S_n(\omega)|$ and phase $\Phi_n(\omega)$ spectra. All SINS spectra here are given by n = 2, i.e., $S_2(\omega)$. For the Far-IR measurements at ALS

Beamline 2.4, a customized Ge:Cu photoconductor, which provides broadband spectral detection down to 320 cm$^{-1}$, and a KRS-5 beamsplitter were employed. Therein, we obtain 20 SINS spectra per point in our measurements, which are averaged and normalized to a reference spectrum obtained from a pure gold surface.

### 4.2 – Calculations

We applied first-principles calculations within density functional theory and density functional perturbation theory[48–50]. FL and 1L-talc were theoretically considered by using a slab model and appropriate boundary conditions for 2D crystals[51,52], with vacuum equal to approximately 16 angstroms. All calculations were performed using Quantum Espresso distribution[53,54]. The structures were optimized until the forces on the atoms and stress on the lattice were lower than 2.57 meV/angstrom and 50 MPa, respectively, with a Brillouin zone sampling by using the Monkhorst-Pack scheme[55], with 4x4x2 and 4x4x1 Gamma center k-point grids for bulk- and FL-talc, respectively. In the linear response calculations[56], to obtain the Raman and IR spectra simulations, the Brillouin zone sampling was calculated by using 6x6x4 and 6x6x1 for bulk and FL-talc, respectively. The valence electron-nucleus interactions were described by using Troullier-Martin norm-conversing pseudopotentials[57], available at Quantum Espresso repositories, setting the kinetic energy cutoff in the wave-function (charge-density) expansion to 46 Eh (184 Eh). We considered the Perdew-Wang's parametrization of the local density approximation for the exchange-correlation (XC) functional[58]. In addition, for bulk, we also considered PBEsol[59] and RVV10[60] XC functionals, where the former is an improved parametrization for solids of the Perdew Burker Ernzerhof PBE XC functional[61] and the latter is an improved implementation of the VV10 XC functional[62], which describes the van der Waals interactions. For estimation of the graphene stiffness coefficients here reported, we built the related stress *vs* strain curves by using the LDA XC functional and Troullier-Martin norm-conserving pseudopotentials, with kinetic energy cutoff in the wave-function (charge-density) expansion equal to 30 Eh (120Eh). The Brillouin zone sampling was performed by using the cold-smearing method, with electronic temperature equal to 15mEh, and 8x8x6 Gamma center within the Monkhorst-Pack method.

**ACKNOWLEDGMENTS**

All Brazilian authors thank the financial support from CAPES, the Advanced Light Source (ALS) for providing beamtime for the experiments and Stephanie N. Gilbert Corder for the experimental assistance, and the LAM support lab from the Brazilian Synchrotron Light Laboratory/CNPEM. NeaSpec GmbH is acknowledged for its technical assistance. R.L., A.R.C., C.J.S.de.M., J.R.-S., and I.D.B. acknowledge the financial support from the Brazilian Nanocarbon Institute of Science and Technology (INCT/Nanocarbono). J.R.-S. and I.D.B. acknowledge the prize L'ORÉAL-UNESCO-ABC for Women in Science Prize – Brazil (2017 and 2021 editions, respectively). J.R.-S. and R.L. thank the support from FAPEMIG through Grants No. APQ-01922-21, APQ-01553-22, RED-00185-16, and RED-00282-16. R.L.M.L. acknowledges the computational time at CENAPAD-SP, CENAPAD-RJ/LNCC, CENAPAD-UFC, SDumont supercomputer. I.D.B., A.R.C, J.R.-S., and R.O.F. acknowledge the CNPq through the research grants 311327/2020-6, 312865/2020-1, 309920/2021-3, 433027/2018-5, 311564/2018-6, 408319/2021-6 and 309946/2021-2. A.R.C. and C.J.S.de.M. acknowledge the FAPESP financial support (Grants No.: 2018/25339-4, 2018/07276-5, and 2020/04374-6) and Fundo Mackenzie de Pesquisa e Inovação (MackPesquisa No. 221017). R.O.F. acknowledges the support from the FAPESP Young Investigator grant 2019/14017-9. This research used resources of the Advanced Light Source, a U.S. DOE Office of Science User Facility under contract No. DE-AC02-05CH11231.

## REFERENCES

(1)     Zhao, W.; Ribeiro, R. M.; Toh, M.; Carvalho, A.; Kloc, C.; Castro Neto, A. H.; Eda, G. Origin of Indirect Optical Transitions in Few-Layer MoS2, WS2, and WSe2. *Nano Lett* **2013**, *13* (11), 5627–5634. https://doi.org/10.1021/nl403270k.

(2)     Longuinhos, R.; Ribeiro-Soares, J. Monitoring the Applied Strain in Monolayer Gallium Selenide through Vibrational Spectroscopies: A First-Principles Investigation. *Phys Rev Appl* **2019**, *11* (2), 24012. https://doi.org/10.1103/PhysRevApplied.11.024012.

(3)     C. Gomes, L.; Carvalho, A. Electronic and Optical Properties of Low-Dimensional Group-IV Monochalcogenides. *J Appl Phys* **2020**, *128* (12), 121101. https://doi.org/10.1063/5.0016003.

(4)     Ribeiro-Soares, J.; Almeida, R. M.; Barros, E. B.; Araujo, P. T.; Dresselhaus, M. S.; Cançado, L. G.; Jorio, A. Group Theory Analysis of Phonons in Two-Dimensional Transition Metal Dichalcogenides. *Phys Rev B* **2014**, *90* (11), 115438. https://doi.org/10.1103/PhysRevB.90.115438.

(5)     Pimenta, M. A.; Corro, E. Del; Carvalho, B. R.; Fantini, C.; Malard, L. M. Comparative Study of Raman Spectroscopy in Graphene and MoS2-Type Transition Metal Dichalcogenides. *Acc Chem Res* **2015**, *48* (1), 41–47. https://doi.org/10.1021/ar500280m.

(6)     Alencar, R. S.; Longuinhos, R.; Rabelo, C.; Miranda, H.; Viana, B. C.; Filho, A. G. S.; Cançado, L. G.; Jorio, A.; Ribeiro-Soares, J. Raman Spectroscopy Polarization Dependence Analysis in Two-Dimensional Gallium Sulfide. *Phys Rev B* **2020**, *102* (16), 1–10. https://doi.org/10.1103/physrevb.102.165307.

(7)     Araujo, F. D. V; Oliveira, V. V; Gadelha, A. C.; Carvalho, T. C. V; Fernandes, T. F. D.; Silva, F. W. N.; Longuinhos, R.; Ribeiro-Soares, J.; Jorio, A.; Souza Filho, A. G.; Alencar, R. S.; Viana, B. C. Temperature-Dependent Phonon Dynamics and Anharmonicity of Suspended and Supported Few-Layer Gallium Sulfide. *Nanotechnology* **2020**, *31* (49), 495702. https://doi.org/10.1088/1361-6528/abb107.

(8)     Longuinhos, R.; Vymazalová, A.; Cabral, A. R.; Ribeiro-Soares, J. Raman Spectrum of Layered Tilkerodeite (Pd 2 HgSe 3 ) Topological Insulator: The Palladium Analogue of Jacutingaite (Pt 2 HgSe 3 ). *Journal of Physics: Condensed Matter* **2021**, *33* (6), 065401. https://doi.org/10.1088/1361-648X/abc35a.

(9)     Ribeiro, H. B.; Ramos, S. L. L. M.; Seixas, L.; de Matos, C. J. S.; Pimenta, M. A. Edge Phonons in Layered Orthorhombic GeS and GeSe Monochalcogenides. *Phys Rev B* **2019**, *100* (9), 94301. https://doi.org/10.1103/PhysRevB.100.094301.

(10)    Barcelos, I. D.; Bechtel, H. A.; de Matos, C. J. S.; Bahamon, D. A.; Kaestner, B.; Maia, F. C. B.; Freitas, R. O. Probing Polaritons in 2D Materials with Synchrotron Infrared Nanospectroscopy. *Adv Opt Mater* **2020**, *8* (5), 1901091. https://doi.org/10.1002/adom.201901091.

(11)    Bechtel, H. A.; Johnson, S. C.; Khatib, O.; Muller, E. A.; Raschke, M. B. Synchrotron Infrared Nano-Spectroscopy and -Imaging. *Surf Sci Rep* **2020**, *75* (3), 100493. https://doi.org/10.1016/j.surfrep.2020.100493.

(12)    Bechtel, H. A.; Muller, E. A.; Olmon, R. L.; Martin, M. C.; Raschke, M. B. Ultrabroadband Infrared Nanospectroscopic Imaging. *Proceedings of the National Academy of Sciences* **2014**, *111* (20), 7191–7196. https://doi.org/10.1073/pnas.1400502111.

(13)    Neal, S. N.; Kim, H.-S.; O'Neal, K. R.; Haglund, A. V.; Smith, K. A.; Mandrus, D. G.; Bechtel, H. A.; Carr, G. L.; Haule, K.; Vanderbilt, D.; Musfeldt, J. L. Symmetry Crossover in Layered MPS3 Complexes (M=Mn, Fe, Ni) via near-Field Infrared Spectroscopy. *Phys Rev B* **2020**, *102* (8), 085408. https://doi.org/10.1103/PhysRevB.102.085408.

(14) Shi, Z.; Bechtel, H. A.; Berweger, S.; Sun, Y.; Zeng, B.; Jin, C.; Chang, H.; Martin, M. C.; Raschke, M. B.; Wang, F. Amplitude- and Phase-Resolved Nanospectral Imaging of Phonon Polaritons in Hexagonal Boron Nitride. *ACS Photonics* **2015**, *2* (7), 790–796. https://doi.org/10.1021/acsphotonics.5b00007.

(15) Neal, S. N.; O'Neal, K. R.; Haglund, A.; Mandrus, D. G.; Bechtel, H.; Carr, G. L.; Haule, K.; Vanderbilt, D.; Kim, H.-S.; Musfeldt, J. L. Exploring Few and Single Layer CrPS 4 with Near-Field Infrared Spectroscopy. *2d Mater* **2021**, 0–12. https://doi.org/10.1088/2053-1583/abf251.

(16) Fali, A.; Gamage, S.; Howard, M.; Folland, T. G.; Mahadik, N. A.; Tiwald, T.; Bolotin, K.; Caldwell, J. D.; Abate, Y. Nanoscale Spectroscopy of Dielectric Properties of Mica. *ACS Photonics* **2021**, *8* (1), 175–181. https://doi.org/10.1021/acsphotonics.0c00951.

(17) Scolfaro, D.; Finamor, M.; Trinchão, L. O.; Rosa, B. L. T.; Chaves, A.; Santos, P. V.; Iikawa, F.; Couto, O. D. D. Acoustically Driven Dynamic Stark Effect in Transition Metal Dichalcogenide Monolayers. *ACS Nano* **2021**, *15* (9), 15371–15380. https://doi.org/10.1021/acsnano.1c06854.

(18) Hamza Safeer, S.; Ore, A. S. M. v; Cadore, A. R.; Gordo, V. O.; Vianna, P. G.; Carvalho, I. C. S.; Carozo, V.; de Matos, C. J. S. CVD Growth and Optical Characterization of Homo and Heterobilayer TMDs. *J Appl Phys* **2022**, *132* (2), 024301. https://doi.org/10.1063/5.0088413.

(19) Peña Román, R. J.; Auad, Y.; Grasso, L.; Padilha, L. A.; Alvarez, F.; Barcelos, I. D.; Kociak, M.; Zagonel, L. F. Design and Implementation of a Device Based on an Off-Axis Parabolic Mirror to Perform Luminescence Experiments in a Scanning Tunneling Microscope. *Review of Scientific Instruments* **2022**, *93* (4), 043704. https://doi.org/10.1063/5.0078423.

(20) Rosa, B. L. T.; Fujisawa, K.; Santos, J. C. C.; Zhang, T.; Matos, M. J. S.; Sousa, F. B.; Barbosa, T. C.; Lafeta, L.; Ramos, S. L. L. M.; Carvalho, B. R.; Chacham, H.; Neves, B. R. A.; Terrones, M.; Malard, L. M. Investigation of Spatially Localized Defects in Synthetic WS2 Monolayers. *Phys Rev B* **2022**, *106* (11), 115301. https://doi.org/10.1103/PhysRevB.106.115301.

(21) Choi, W.; Choudhary, N.; Han, G. H.; Park, J.; Akinwande, D.; Lee, Y. H. Recent Development of Two-Dimensional Transition Metal Dichalcogenides and Their Applications. *Materials Today* **2017**, *20* (3), 116–130. https://doi.org/10.1016/j.mattod.2016.10.002.

(22) Backes, C.; Abdelkader, A. M.; Alonso, C.; Andrieux-Ledier, A.; Arenal, R.; Azpeitia, J.; Balakrishnan, N.; Banszerus, L.; Barjon, J.; Bartali, R.; Bellani, S.; Berger, C.; Berger, R.; Ortega, M. M. B.; Bernard, C.; Beton, P. H.; Beyer, A.; Bianco, A.; Bøggild, P.; Bonaccorso, F.; Barin, G. B.; Botas, C.; Bueno, R. A.; Carriazo, D.; Castellanos-Gomez, A.; Christian, M.; Ciesielski, A.; Ciuk, T.; Cole, M. T.; Coleman, J.; Coletti, C.; Crema, L.; Cun, H.; Dasler, D.; De Fazio, D.; Díez, N.; Drieschner, S.; Duesberg, G. S.; Fasel, R.; Feng, X.; Fina, A.; Forti, S.; Galiotis, C.; Garberoglio, G.; García, J. M.; Garrido, J. A.; Gibertini, M.; Gölzhäuser, A.; Gómez, J.; Greber, T.; Hauke, F.; Hemmi, A.; Hernandez-Rodriguez, I.; Hirsch, A.; Hodge, S. A.; Huttel, Y.; Jepsen, P. U.; Jimenez, I.; Kaiser, U.; Kaplas, T.; Kim, H.; Kis, A.; Papagelis, K.; Kostarelos, K.; Krajewska, A.; Lee, K.; Li, C.; Lipsanen, H.; Liscio, A.; Lohe, M. R.; Loiseau, A.; Lombardi, L.; Francisca López, M.; Martin, O.; Martín, C.; Martínez, L.; Martin-Gago, J. A.; Ignacio Martínez, J.; Marzari, N.; Mayoral, Á.; McManus, M.; Melucci, M.; Méndez, J.; Merino, C.; Merino, P.; Meyer, A. P.; Miniussi, E.; Miseikis, V.; Mishra, N.; Morandi, V.; Munuera, C.; Muñoz, R.; Nolan, H.; Ortolani, L.; Ott, A. K.; Palacio, I.; Palermo, V.; Parthenios, J.; Pasternak, I.; Patane, A.; Prato, M.; Prevost, H.; Prudkovskiy, V.; Pugno, N.; Rojo, T.; Rossi, A.; Ruffieux, P.; Samorì, P.; Schué, L.; Setijadi, E.; Seyller, T.; Speranza, G.; Stampfer, C.; Stenger, I.; Strupinski, W.; Svirko, Y.; Taioli, S.; Teo, K. B. K.; Testi, M.; Tomarchio, F.; Tortello, M.; Treossi, E.; Turchanin, A.; Vazquez, E.; Villaro, E.; Whelan, P. R.; Xia, Z.; Yakimova, R.; Yang, S.; Yazdi, G. R.; Yim, C.; Yoon, D.; Zhang, X.; Zhuang, X.; Colombo, L.; Ferrari, A. C.; Garcia-

Hernandez, M. Production and Processing of Graphene and Related Materials. *2d Mater* **2020**, *7* (2), 022001. https://doi.org/10.1088/2053-1583/ab1e0a.

(23)   Frisenda, R.; Niu, Y.; Gant, P.; Muñoz, M.; Castellanos-Gomez, A. Naturally Occurring van Der Waals Materials. *NPJ 2D Mater Appl* **2020**, *4* (1), 1–13. https://doi.org/10.1038/s41699-020-00172-2.

(24)   Santos, J. C. C.; Barboza, A. P. M.; Matos, M. J. S.; Barcelos, I. D.; Fernandes, T. F. D.; Soares, E. A.; Moreira, R. L.; Neves, B. R. A. Exfoliation and Characterization of a Two-Dimensional Serpentine-Based Material. *Nanotechnology* **2019**, *30* (44), 445705. https://doi.org/10.1088/1361-6528/ab3732.

(25)   de Oliveira, R.; Guallichico, L. A. G.; Policarpo, E.; Cadore, A. R.; Freitas, R. O.; da Silva, F. M. C.; de C. Teixeira, V.; Paniago, R. M.; Chacham, H.; Matos, M. J. S.; Malachias, A.; Krambrock, K.; Barcelos, I. D. High Throughput Investigation of an Emergent and Naturally Abundant 2D Material: Clinochlore. *Appl Surf Sci* **2022**, *599* (December 2021), 153959. https://doi.org/10.1016/j.apsusc.2022.153959.

(26)   Cadore, A. R.; de Oliveira, R.; Longuinhos, R.; Teixeira, V. de C.; Nagaoka, D. A.; Alvarenga, V. T.; Ribeiro-Soares, J.; Watanabe, K.; Taniguchi, T.; Paniago, R. M.; Malachias, A.; Krambrock, K.; Barcelos, I. D.; de Matos, C. J. S. Exploring the Structural and Optoelectronic Properties of Natural Insulating Phlogopite in van Der Waals Heterostructures. *2d Mater* **2022**, *9* (3). https://doi.org/10.1088/2053-1583/ac6cf4.

(27)   Batista, R. J. C.; Dias, R. F.; Barboza, A. P. M.; de Oliveira, A. B.; Manhabosco, T. M.; Gomes-Silva, T. R.; Matos, M. J. S.; Gadelha, A. C.; Rabelo, C.; Cançado, L. G. L.; Jorio, A.; Chacham, H.; Neves, B. R. A. Nanomechanics of Few-Layer Materials: Do Individual Layers Slide upon Folding? *Beilstein Journal of Nanotechnology* **2020**, *11*, 1801–1808. https://doi.org/10.3762/bjnano.11.162.

(28)   Mania, E.; Alencar, A. B.; Cadore, A. R.; Carvalho, B. R.; Watanabe, K.; Taniguchi, T.; Neves, B. R. A.; Chacham, H.; Campos, L. C. Spontaneous Doping on High Quality Talc-Graphene-HBN van Der Waals Heterostructures. *2d Mater* **2017**, *4* (3), 031008. https://doi.org/10.1088/2053-1583/aa76f4.

(29)   Nutting, D.; Prando, G. A.; Severijnen, M.; Barcelos, I. D.; Guo, S.; Christianen, P. C. M.; Zeitler, U.; Galvão Gobato, Y.; Withers, F. Electrical and Optical Properties of Transition Metal Dichalcogenides on Talc Dielectrics. *Nanoscale* **2021**, *13* (37), 15853–15858. https://doi.org/10.1039/D1NR04723J.

(30)   Farmer, V. C.; Russell, J. D. The Infra-Red Spectra of Layer Silicates. *Spectrochimica Acta* **1964**, *20* (7), 1149–1173. https://doi.org/10.1016/0371-1951(64)80165-X.

(31)   Farmer, V. C. The Infra-Red Spectra of Talc, Saponite, and Hectorite. *Mineralogical Magazine and Journal of the Mineralogical Society* **1958**, *31* (241), 829–845. https://doi.org/10.1180/minmag.1958.031.241.03.

(32)   Tufar, W. Talc. In *Ullmann's Encyclopedia of Industrial Chemistry*; Wiley-VCH Verlag GmbH & Co. KGaA: Weinheim, Germany, 2000. https://doi.org/10.1002/14356007.a26_047.

(33)   Matković, A.; Ludescher, L.; Peil, O. E.; Sharma, A.; Gradwohl, K.-P.; Kratzer, M.; Zimmermann, M.; Genser, J.; Knez, D.; Fisslthaler, E.; Gammer, C.; Lugstein, A.; Bakker, R. J.; Romaner, L.; Zahn, D. R. T.; Hofer, F.; Salvan, G.; Raith, J. G.; Teichert, C. Iron-Rich Talc as Air-Stable Platform for Magnetic Two-Dimensional Materials. *NPJ 2D Mater Appl* **2021**, *5* (1), 94. https://doi.org/10.1038/s41699-021-00276-3.

(34)   Barcelos, I. D.; Cadore, A. R.; Alencar, A. B.; Maia, F. C. B.; Mania, E.; Oliveira, R. F.; Bufon, C. C. B.; Malachias, Â.; Freitas, R. O.; Moreira, R. L.; Chacham, H. Infrared Fingerprints of Natural 2D Talc and Plasmon–Phonon Coupling in Graphene–Talc Heterostructures. *ACS Photonics* **2018**, *5* (5), 1912–1918. https://doi.org/10.1021/acsphotonics.7b01017.

(35) Gadelha, A. C.; Vasconcelos, T. L.; Cançado, L. G.; Jorio, A. Nano-Optical Imaging of In-Plane Homojunctions in Graphene and MoS 2 van Der Waals Heterostructures on Talc and SiO 2. *J Phys Chem Lett* **2021**, *12* (31), 7625–7631. https://doi.org/10.1021/acs.jpclett.1c01804.

(36) Prando, G. A.; Severijnen, M. E.; Barcelos, I. D.; Zeitler, U.; Christianen, P. C. M.; Withers, F.; Galvão Gobato, Y. Revealing Excitonic Complexes in Monolayer ${\mathrm{WS}}_{2}$ on Talc Dielectric. *Phys Rev Appl* **2021**, *16* (6), 064055. https://doi.org/10.1103/PhysRevApplied.16.064055.

(37) Cadore, A. R.; Mania, E.; Alencar, A. B.; Rezende, N. P.; de Oliveira, S.; Watanabe, K.; Taniguchi, T.; Chacham, H.; Campos, L. C.; Lacerda, R. G. Enhancing the Response of NH3 Graphene-Sensors by Using Devices with Different Graphene-Substrate Distances. *Sens Actuators B Chem* **2018**, *266*, 438–446. https://doi.org/10.1016/j.snb.2018.03.164.

(38) Kirak, A.; Yilmaz, H.; Güler, S.; Güler, Ç. Dielectric Properties and Electric Conductivity of Talc and Doped Talc. *J Phys D Appl Phys* **1999**, *32* (15), 1919–1927. https://doi.org/10.1088/0022-3727/32/15/321.

(39) Alencar, A. B.; Barboza, A. P. M.; Archanjo, B. S.; Chacham, H.; Neves, B. R. A. Experimental and Theoretical Investigations of Monolayer and Fewlayer Talc. *2d Mater* **2015**, *2* (1), 015004. https://doi.org/10.1088/2053-1583/2/1/015004.

(40) Chacham, H.; Barboza, A. P. M.; de Oliveira, A. B.; de Oliveira, C. K.; Batista, R. J. C.; Neves, B. R. A. Universal Deformation Pathways and Flexural Hardening of Nanoscale 2D-Material Standing Folds. *Nanotechnology* **2018**, *29* (9), 095704. https://doi.org/10.1088/1361-6528/aaa51e.

(41) Batista, R. J. C.; Dias, R. F.; Barboza, A. P. M.; de Oliveira, A. B.; Manhabosco, T. M.; Gomes-Silva, T. R.; Matos, M. J. S.; Gadelha, A. C.; Rabelo, C.; Cançado, L. G. L.; Jorio, A.; Chacham, H.; Neves, B. R. A. Nanomechanics of Few-Layer Materials: Do Individual Layers Slide upon Folding? *Beilstein Journal of Nanotechnology* **2020**, *11*, 1801–1808. https://doi.org/10.3762/bjnano.11.162.

(42) Claverie, M.; Dumas, A.; Carême, C.; Poirier, M.; Le Roux, C.; Micoud, P.; Martin, F.; Aymonier, C. Synthetic Talc and Talc-Like Structures: Preparation, Features and Applications. *Chemistry - A European Journal* **2018**, *24* (3), 519–542. https://doi.org/10.1002/chem.201702763.

(43) Vasic, B.; Czibula, C.; Kratzer, M.; Neves, B. R. A.; Matkovic, A.; Teichert, C. Two-Dimensional Talc as a van Der Waals Material for Solid Lubrication at the Nanoscale. *Nanotechnology* **2021**, *127* (1), 69–73. https://doi.org/10.1088/1361-6528/abeffe.

(44) Feres, F. H.; Mayer, R. A.; Wehmeier, L.; Maia, F. C. B.; Viana, E. R.; Malachias, A.; Bechtel, H. A.; Klopf, J. M.; Eng, L. M.; Kehr, S. C.; González, J. C.; Freitas, R. O.; Barcelos, I. D. Sub-Diffractional Cavity Modes of Terahertz Hyperbolic Phonon Polaritons in Tin Oxide. *Nat Commun* **2021**, *12* (1), 1995. https://doi.org/10.1038/s41467-021-22209-w.

(45) Barcelos, I. D.; Canassa, T. A.; Mayer, R. A.; Feres, F. H.; de Oliveira, E. G.; Goncalves, A.-M. B.; Bechtel, H. A.; Freitas, R. O.; Maia, F. C. B.; Alves, D. C. B. Ultrabroadband Nanocavity of Hyperbolic Phonon–Polaritons in 1D-Like α-MoO 3. *ACS Photonics* **2021**, acsphotonics.1c00955. https://doi.org/10.1021/acsphotonics.1c00955.

(46) So, S.; Kim, M.; Lee, D.; Nguyen, D. M.; Rho, J. Overcoming Diffraction Limit: From Microscopy to Nanoscopy. *Appl Spectrosc Rev* **2018**, *53* (2–4), 290–312. https://doi.org/10.1080/05704928.2017.1323309.

(47) Ardito, F. M.; Mendes-De-Sá, T. G.; Cadore, A. R.; Gomes, P. F.; Mafra, D. L.; Barcelos, I. D.; Lacerda, R. G.; Iikawa, F.; Granado, E. Damping of Landau Levels in Neutral Graphene at Low Magnetic Fields: A Phonon Raman Scattering Study. *Phys Rev B* **2018**, *97* (3), 1–6. https://doi.org/10.1103/PhysRevB.97.035419.

(48) Hohenberg, P.; Kohn, W. Inhomogeneous Electron Gas. *Physical Review* **1964**, *136* (3B), B864–B871.

(49) Kohn, W.; Sham, L. J. Self-Consistent Equations Including Exchange and Correlation Effects. *Physical Review* **1965**, *140* (4A), A1133–A1138.

(50) Baroni, S.; de Gironcoli, S.; Dal Corso, A.; Giannozzi, P. Phonons and Related Crystal Properties from Density-Functional Perturbation Theory. *Rev Mod Phys* **2001**, *73* (2), 515–562.

(51) Sohier, T.; Calandra, M.; Mauri, F. Density Functional Perturbation Theory for Gated Two-Dimensional Heterostructures: Theoretical Developments and Application to Flexural Phonons in Graphene. *Phys Rev B* **2017**, *96* (7), 075448. https://doi.org/10.1103/PhysRevB.96.075448.

(52) Sohier, T.; Gibertini, M.; Calandra, M.; Mauri, F.; Marzari, N. Breakdown of Optical Phonons' Splitting in Two-Dimensional Materials. *Nano Lett* **2017**, *17* (6), 3758–3763. https://doi.org/10.1021/acs.nanolett.7b01090.

(53) Giannozzi, P.; Baroni, S.; Bonini, N.; Calandra, M.; Car, R.; Cavazzoni, C.; Ceresoli, D.; Chiarotti, G. L.; Cococcioni, M.; Dabo, I.; Dal Corso, A.; de Gironcoli, S.; Fabris, S.; Fratesi, G.; Gebauer, R.; Gerstmann, U.; Gougoussis, C.; Kokalj, A.; Lazzeri, M.; Martin-Samos, L.; Marzari, N.; Mauri, F.; Mazzarello, R.; Paolini, S.; Pasquarello, A.; Paulatto, L.; Sbraccia, C.; Scandolo, S.; Sclauzero, G.; Seitsonen, A. P.; Smogunov, A.; Umari, P.; Wentzcovitch, R. M. QUANTUM ESPRESSO: A Modular and Open-Source Software Project for Quantum Simulations of Materials. *Journal of Physics: Condensed Matter* **2009**, *21* (39), 395502. https://doi.org/10.1088/0953-8984/21/39/395502.

(54) Giannozzi, P.; Andreussi, O.; Brumme, T.; Bunau, O.; Buongiorno Nardelli, M.; Calandra, M.; Car, R.; Cavazzoni, C.; Ceresoli, D.; Cococcioni, M.; Colonna, N.; Carnimeo, I.; Dal Corso, A.; de Gironcoli, S.; Delugas, P.; DiStasio, R. A.; Ferretti, A.; Floris, A.; Fratesi, G.; Fugallo, G.; Gebauer, R.; Gerstmann, U.; Giustino, F.; Gorni, T.; Jia, J.; Kawamura, M.; Ko, H.-Y.; Kokalj, A.; Küçükbenli, E.; Lazzeri, M.; Marsili, M.; Marzari, N.; Mauri, F.; Nguyen, N. L.; Nguyen, H.-V.; Otero-de-la-Roza, A.; Paulatto, L.; Poncé, S.; Rocca, D.; Sabatini, R.; Santra, B.; Schlipf, M.; Seitsonen, A. P.; Smogunov, A.; Timrov, I.; Thonhauser, T.; Umari, P.; Vast, N.; Wu, X.; Baroni, S. Advanced Capabilities for Materials Modelling with Quantum ESPRESSO. *Journal of Physics: Condensed Matter* **2017**, *29* (46), 465901. https://doi.org/10.1088/1361-648X/aa8f79.

(55) Pack, J. D.; Monkhorst, H. J. Special Points for Brillouin-Zone Integrations. *Phys Rev B* **1977**, *16* (4), 1748–1749.

(56) Lazzeri, M.; Mauri, F. First-Principles Calculation of Vibrational Raman Spectra in Large Systems: Signature of Small Rings in Crystalline $SiO_2$. *Phys Rev Lett* **2003**, *90* (3), 036401. https://doi.org/10.1103/PhysRevLett.90.036401.

(57) Troullier, N.; Martins, J. L. Efficient Pseudopotentials for Plane-Wave Calculations. *Phys Rev B* **1991**, *43* (3), 1993. https://doi.org/10.1103/PhysRevB.43.8861.

(58) Perdew, J. P.; Wang, Y. Accurate and Simple Analytic Representation of the Electron-Gas Correlation Energy. *Phys Rev B* **1992**, *45* (23), 13244–13249. https://doi.org/10.1103/PhysRevB.45.13244.

(59) Perdew, J. P.; Ruzsinszky, A.; Csonka, G. I.; Vydrov, O. A.; Scuseria, G. E.; Constantin, L. A.; Zhou, X.; Burke, K. Restoring the Density-Gradient Expansion for Exchange in Solids and Surfaces. *Phys Rev Lett* **2008**, *100* (13), 136406. https://doi.org/10.1103/PhysRevLett.100.136406.

(60) Sabatini, R.; Gorni, T.; de Gironcoli, S. Nonlocal van Der Waals Density Functional Made Simple and Efficient. *Phys Rev B* **2013**, *87* (4), 41108. https://doi.org/10.1103/PhysRevB.87.041108.

(61) Perdew, J. P.; Burke, K.; Ernzerhof, M. Generalized Gradient Approximation Made Simple. *Phys Rev Lett* **1996**, *77* (18), 3865–3868.

(62) Vydrov, O. A.; Van Voorhis, T. Nonlocal van Der Waals Density Functional: The Simpler the Better. *J Chem Phys* **2010**, *133* (24), 244103. https://doi.org/10.1063/1.3521275.

(63) TAYLOR, D. G.; NENADIC, C. M.; CRABLE, J. v. Infrared Spectra for Mineral Identification. *Am Ind Hyg Assoc J* **1970**, *31* (1), 100–108. https://doi.org/10.1080/0002889708506215.

(64) Muller, E. A.; Pollard, B.; Bechtel, H. A.; van Blerkom, P.; Raschke, M. B. Infrared Vibrational Nanocrystallography and Nanoimaging. *Sci Adv* **2016**, *2* (10), e1601006–e1601006. https://doi.org/10.1126/sciadv.1601006.

(65) Neal, S. N.; O'Neal, K. R.; Haglund, A.; Mandrus, D. G.; Bechtel, H.; Carr, G. L.; Haule, K.; Vanderbilt, D.; Kim, H.-S.; Musfeldt, J. L. Exploring Few and Single Layer CrPS 4 with Near-Field Infrared Spectroscopy. *2d Mater* **2021**, 0–12. https://doi.org/10.1088/2053-1583/abf251.

(66) Molina-Sánchez, A.; Wirtz, L. Phonons in Single-Layer and Few-Layer MoS$_{2}$ and WS$_{2}$. *Phys Rev B* **2011**, *84* (15), 155413. https://doi.org/10.1103/PhysRevB.84.155413.

(67) Alencar, R. S.; Longuinhos, R.; Rabelo, C.; Miranda, H.; Viana, B. C.; Filho, A. G. S.; Cançado, L. G.; Jorio, A.; Ribeiro-Soares, J. Raman Spectroscopy Polarization Dependence Analysis in Two-Dimensional Gallium Sulfide. *Phys Rev B* **2020**, *102* (16), 1–10. https://doi.org/10.1103/physrevb.102.165307.

(68) Kuroda, N.; Nishina, Y. Davydov Splitting of Degenerate Lattice Modes in the Layer Compound GaS. *Phys Rev B* **1979**, *19* (2), 1312–1315. https://doi.org/10.1103/PhysRevB.19.1312.

(69) Wang, A.; Freeman, J. J.; Jolliff, B. L. Understanding the Raman Spectral Features of Phyllosilicates. *Journal of Raman Spectroscopy* **2015**, *46* (10), 829–845. https://doi.org/10.1002/jrs.4680.

(70) Petit, S.; Martin, F.; Wiewiora, A.; De Parseval, P.; Decarreau, A. Crystal-Chemistry of Talc: A near Infrared (NIR) Spectroscopy Study. *American Mineralogist* **2004**, *89* (2–3), 319–326. https://doi.org/10.2138/am-2004-2-310.

(71) Matković, A.; Ludescher, L.; Peil, O. E.; Sharma, A.; Gradwohl, K.-P.; Kratzer, M.; Zimmermann, M.; Genser, J.; Knez, D.; Fisslthaler, E.; Gammer, C.; Lugstein, A.; Bakker, R. J.; Romaner, L.; Zahn, D. R. T.; Hofer, F.; Salvan, G.; Raith, J. G.; Teichert, C. Iron-Rich Talc as Air-Stable Platform for Magnetic Two-Dimensional Materials. *NPJ 2D Mater Appl* **2021**, *5* (1), 94. https://doi.org/10.1038/s41699-021-00276-3.

(72) Loh, E. Optical Vibrations in Sheet Silicates. *Journal of Physics C: Solid State Physics* **1973**, *6* (6), 1091–1104. https://doi.org/10.1088/0022-3719/6/6/022.

(73) Reynard, B.; Bezacier, L.; Caracas, R. Serpentines, Talc, Chlorites, and Their High-Pressure Phase Transitions: A Raman Spectroscopic Study. *Phys Chem Miner* **2015**, *42* (8), 641–649. https://doi.org/10.1007/s00269-015-0750-0.

(74) Rosasco, G. J.; Blaha, J. J. Raman Microprobe Spectra and Vibrational Mode Assignments of Talc. *Appl Spectrosc* **1980**, *34* (2), 140–144. https://doi.org/10.1366/0003702804730664.

(75) Blaha, J. J.; Rosasco, G. J. Raman Microprobe Spectra of Individual Microcrystals and Fibers of Talc, Tremolite, and Related Silicate Minerals. *Anal Chem* **1978**, *50* (7), 892–896. https://doi.org/10.1021/ac50029a018.

(76) Alencar, R. S.; Longuinhos, R.; Rabelo, C.; Miranda, H.; Viana, B. C.; Filho, A. G. S.; Cançado, L. G.; Jorio, A.; Ribeiro-Soares, J. Raman Spectroscopy Polarization Dependence Analysis in Two-Dimensional Gallium Sulfide. *Phys Rev B* **2020**, *102* (16), 165307. https://doi.org/10.1103/PhysRevB.102.165307.

(77) Michel, K. H.; Verberck, B. Theory of Elastic and Piezoelectric Effects in Two-Dimensional Hexagonal Boron Nitride. *Phys Rev B* **2009**, *80* (22), 224301. https://doi.org/10.1103/PhysRevB.80.224301.

(78) Michel, K. H.; Verberck, B. Theory of Rigid-Plane Phonon Modes in Layered Crystals. *Phys Rev B* **2012**, *85* (9), 094303. https://doi.org/10.1103/PhysRevB.85.094303.

(79) Tan, P. H.; Han, W. P.; Zhao, W. J.; Wu, Z. H.; Chang, K.; Wang, H.; Wang, Y. F.; Bonini, N.; Marzari, N.; Pugno, N.; Savini, G.; Lombardo, A.; Ferrari, A. C. The Shear Mode of Multilayer Graphene. *Nat Mater* **2012**, *11* (4), 294–300. https://doi.org/10.1038/nmat3245.

(80) Longuinhos, R.; Ribeiro-Soares, J. Ultra-Weak Interlayer Coupling in Two-Dimensional Gallium Selenide. *Physical Chemistry Chemical Physics* **2016**, *18* (36), 25401–25408. https://doi.org/10.1039/C6CP03806A.

(81) Pizzi, G.; Milana, S.; Ferrari, A. C.; Marzari, N.; Gibertini, M. Shear and Breathing Modes of Layered Materials. *ACS Nano* **2021**, *15* (8), 12509–12534. https://doi.org/10.1021/acsnano.0c10672.

(82) He, J.; Paradisanos, I.; Liu, T.; Cadore, A. R.; Liu, J.; Churaev, M.; Wang, R. N.; Raja, A. S.; Javerzac-Galy, C.; Roelli, P.; Fazio, D. de; Rosa, B. L. T.; Tongay, S.; Soavi, G.; Ferrari, A. C.; Kippenberg, T. J. Low-Loss Integrated Nanophotonic Circuits with Layered Semiconductor Materials. *Nano Lett* **2021**, *21* (7), 2709–2718. https://doi.org/10.1021/acs.nanolett.0c04149.

(83) Longuinhos, R.; Vymazalová, A.; Cabral, A. R.; Alexandre, S. S.; Nunes, R. W.; Ribeiro-Soares, J. Raman Spectrum of Layered Jacutingaite (Pt2HgSe3) Crystals—Experimental and Theoretical Study. *Journal of Raman Spectroscopy* **2020**, *51* (2), 357–365. https://doi.org/10.1002/jrs.5764.

(84) Longuinhos Monteiro Lobato, R.; Ribeiro-Soares, J. Mechanical Properties of Layered Tilkerodeite (Pd 2 HgSe 3 ) and Jacutingaite (Pt 2 HgSe 3 ) Crystals: Insights on the Interlayer, Intralayer Interactions, and Phonons. *J Appl Phys* **2021**, *130* (1), 015105. https://doi.org/10.1063/5.0053171.

(85) Tan, P. H.; Han, W. P.; Zhao, W. J.; Wu, Z. H.; Chang, K.; Wang, H.; Wang, Y. F.; Bonini, N.; Marzari, N.; Pugno, N.; Savini, G.; Lombardo, A.; Ferrari, A. C. The Shear Mode of Multilayer Graphene. *Nat Mater* **2012**, *11* (4), 294–300. https://doi.org/10.1038/nmat3245.

# Supporting information

**Raman and Far Infrared Synchrotron Nanospectroscopy of Layered Crystalline Talc:**

**Vibrational Properties, Interlayer Coupling and Symmetry Crossover**

Raphael Longuinhos[1,7], Alisson R. Cadore[2,3,7], Hans A. Bechtel[4], Christiano J. S. de Matos[3,5], Raul O. Freitas[6], Jenaina Ribeiro-Soares[1+], Ingrid D. Barcelos[6+]

[1]*Departamento de Física, Universidade Federal de Lavras (UFLA)*, Zip Code 37200-900, Lavras, Minas Gerais, Brazil*.*
[2]*Present address: Brazilian Nanotechnology National Laboratory (LNNano), Brazilian Center for Research in Energy and Materials (CNPEM), Zip Code 13083-970, Campinas, Sao Paulo, Brazil.*
[3]*School of Engineering, Mackenzie Presbyterian University, Zip Code 01302-907, São Paulo, Brazil.*
[4]*Advanced Light Source (ALS), Lawrence Berkeley National Laboratory, Zip Code 94720, Berkeley, California, USA.*
[5]*MackGraphe, Mackenzie Presbyterian Institute, Zip Code 01302-907, São Paulo, Brazil.*
[6]*Brazilian Synchrotron Light Laboratory (LNLS), Brazilian Center for Research in Energy and Materials (CNPEM), Zip Code 13083-970, Campinas, Sao Paulo, Brazil.*
[7]These authors contributed equally to this work.
+Corresponding-Author: jenaina.soares@ufla.br, ingrid.barcelos@lnls.br

## Simulations of the structure and Raman-active lattice vibrations of bulk talc

In Table S1, we present the results for bulk-talc structural parameters, as calculated by using LDA [1], PBEsol [2,3] and rVV10 [4,5] exchange-correlation (XC) functionals. The calculations were performed with the same parameters described in the main text, but the kinetic energy cutoff for the wavefunctions (charge density) in the rVV10 calculations, equal to 49 $E_h$ (196 $E_h$). All the XC display excellent agreement to the experimental values from this work and Reference [6] (we included the results from the latter in Table 1). In more details, considering the lattice parameters a, b, and c, we analyzed the mean relative error $MRE = \frac{1}{N}\sum_{i=1}^{N}\frac{x_i-y_i}{y_i}$ and mean $MARE = \frac{1}{N}\sum_{i}^{N}\left|\frac{x_i-y_i}{y_i}\right|$, $x_i$ and $y_i$ are the simulated and measured values, respectively. We notice that the van der Waals functional rVV10 displays the best agreement to experiments. LDA, which does not account explicitly the van der Waals interactions, displays good agreement to experiments, attributed to error cancellation[7]The MRE and MARE show that LDA systematically underestimates the experimental values, while PBEsol systematically overestimates the experimental values. The PBEsol improves the in-plane lattice description in comparison to LDA, and rVV10 improves both in-plane and stacking lattice parameters in comparison to both LDA and PBEsol.

*Table S1: Bulk talc structural parameters, as calculated by using LDA, PBEsol and rVV10 XC functionals, and measured by using X-ray [8]. Lattice constants a, b, and c are in units of Å  lattice volume is in units of Å³. Lattice angles, α, β, and γ are in degrees.*

|  | Ref. [7] | LDA | PBEsol | rVV10 | rVV10 |
|---|---|---|---|---|---|
| a | 5.291 | 5.211 | 5.301 | 5.302 | 5.302 |
| b | 5.290 | 5.211 | 5.301 | 5.302 | 5.302 |
| c | 9.460 | 9.187 | 9.817 | 9.391 | 9.4 |
| MRE (%) |  | -1.96 | 1.39 | -0.10 | 0.07 |
| MARE(%) |  | 1.96 | 1.39 | 0.39 | 0.36 |
|  |  |  |  |  |  |
| α | 81.32 | 79.94 | 80.38 | 79.89 | 79.75 |
| β | 85.27 | 84.17 | 84.38 | 84.20 | 84.05 |
| γ | 60.10 | 60.16 | 60.03 | 60.15 | 61.15 |
| Volume | 226.90 | 213.04 | 235.63 | 225.39 | 225.48 |

In Table S2, we present the results for the wavenumbers of the bulk talc first-order Raman active modes, as calculated by using the aforementioned XC functionals. For the MRE and MARE analysis, we considered the experimental values in this work. We considered only one simulated value (in bold in the Table) when more than one is related to a single experimental value, and the average among experimental values related to a single simulated value. We excluded cases where no experimental value was detected. The results show that in the case of wavenumber calculations, LDA outperforms both PBEsol and rVV10 XC functionals.

Table S2: Symmetry assignment (Irr) and wavenumber (in cm$^{-1}$) of the Raman-active modes in bulk talc, as calculated by using LDA, PBEsol and rVV10 exchange-correlation functionals. The assignment of the experimental modes from this work or from literature is tempting for some cases, written in bold, as there are many candidates within a narrow wavenumber range.

| Irr. | ω LDA | ω PBEsol | ω rVV10 | ω Exp. [this work] | ω Exp. Ref. [9] | ω Exp. Ref. [10] |
|---|---|---|---|---|---|---|
| $A_g$ | 106.9 | 86.3 | 106.3 | 104.0 | | 109.0 |
| $A_g$ | 119.4 | 101.4 | 115.5 | 111.0/116.0 | | 113.0/119.0 |
| $A_g$ | 133.1 | 104.5 | 121.6 | | | |
| $A_g$ | 198.7 | 180.2 | 198.0 | 195.0 | 193.0 | 197.0 |
| $A_g$ | 228.0 | 219.0 | 237.1 | 229.0 | | 232.0 |
| $A_g$ | 288.6 | 276.7 | 288.1 | 292.0 | | 294.0 |
| $A_g$ | 303.6 | 290.7 | 301.9 | 292.0 | | 307.0 |
| $A_g$ | **333.9** | **320.1** | **325.0** | 333.0 | | 335.0 |
| $A_g$ | 338.0 | 325.4 | 328.3 | | | |
| $A_g$ | 348.1 | 334.7 | 339.0 | | | |
| $A_g$ | 366.3 | 352.3 | 350.2 | 362.0 | 361.0 | 366.0 |
| $A_g$ | 374.4 | 365.3 | 370.1 | | | |
| $A_g$ | 385.4 | 372.8 | 380.1 | 381.0 | | 383.0 |
| $A_g$ | 428.2 | 415.0 | 422.6 | 432.0 | | 435.0 |
| $A_g$ | **430.5** | **416.3** | **422.9** | | | |
| $A_g$ | 456.1 | 440.1 | 444.5 | 452.0 | | 456.0 |
| $A_g$ | 473.4 | 457.1 | 460.8 | 468.0 | | 471.0 |
| $A_g$ | 506.2 | 485.9 | 498.3 | 508.0 | | 511.0 |
| $A_g$ | 517.0 | 497.2 | 509.2 | 517.0 | | 519.0 |
| $A_g$ | 658.7 | 647.3 | 655.6 | 677.0 | 675.0 | 679.0 |
| $A_g$ | 661.9 | 648.2 | 673.8 | | | |
| $A_g$ | **668.8** | **653.0** | **676.7** | | | |
| $A_g$ | 788.9 | 772.5 | 758.4 | 785.0 | | 789.0 |
| $A_g$ | 795.9 | 777.9 | 766.0 | 792.0 | | 795.0 |
| $A_g$ | 906.6 | 889.6 | 870.6 | | | |
| $A_g$ | 1017.5 | 1003.5 | 969.9 | 1015.0 | | 1018.0 |
| $A_g$ | 1023.4 | 1004.1 | 976.5 | | | |
| $A_g$ | 1047.7 | 1030.0 | 1013.4 | 1051.0 | 1051.0 | 1049.0 |
| $A_g$ | 1084.3 | 1073.9 | 1050.1 | | | |
| $A_g$ | 3667.6 | 3736.0 | 3706.8 | 3663.0/3679.0 | 3663.0/3678.0 | 3675.0 |
| MRE(%) | 0.67 | -4.09 | -1.04 | | | |
| MARE(%) | 1.05 | 4.28 | 2.16 | | | |

**Experimental and calculated infrared response dependence to the number of layers**

In Figure S1, we display the simulated IR spectra of powder layered talc (using proper 2D periodic boundary [11]). We found negligible wavenumber dependence with N-layers.

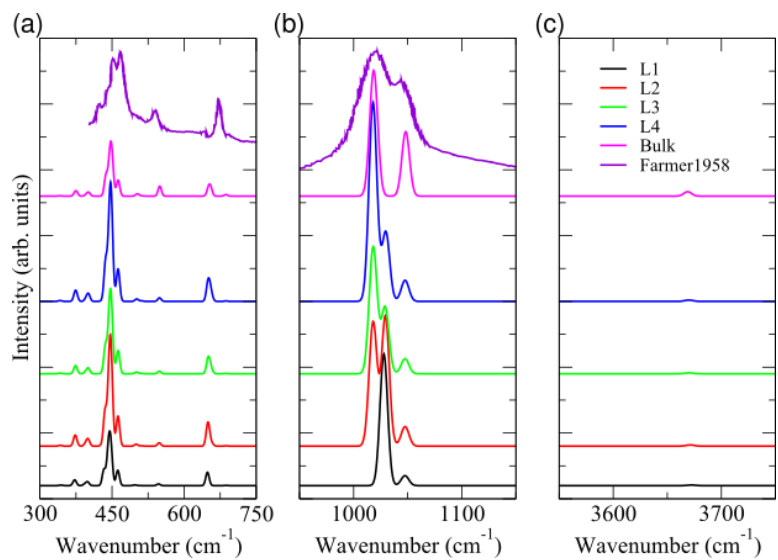

Figure S1: Trends of the IR spectra of powder layered talc, from 1L to bulk limit. Solid lines are Gaussian convolution of the relative intensities of the infrared-active modes. Experimental results from Reference [12].

Figure S2 displays the effects of the dimensionality on the wavenumber of the IR-active modes in layered talc for the peaks at 437.6 cm$^{-1}$, 502.6 cm$^{-1}$, 549.3 cm$^{-1}$ and 686.95 cm$^{-1}$, where the N-layer wavenumbers represent the intensity-weighted average of the wavenumber of the modes observed at the vicinity of the bulk modes and we followed the atomic vibration pattern of the bulk mode down to the FL-talc regime.

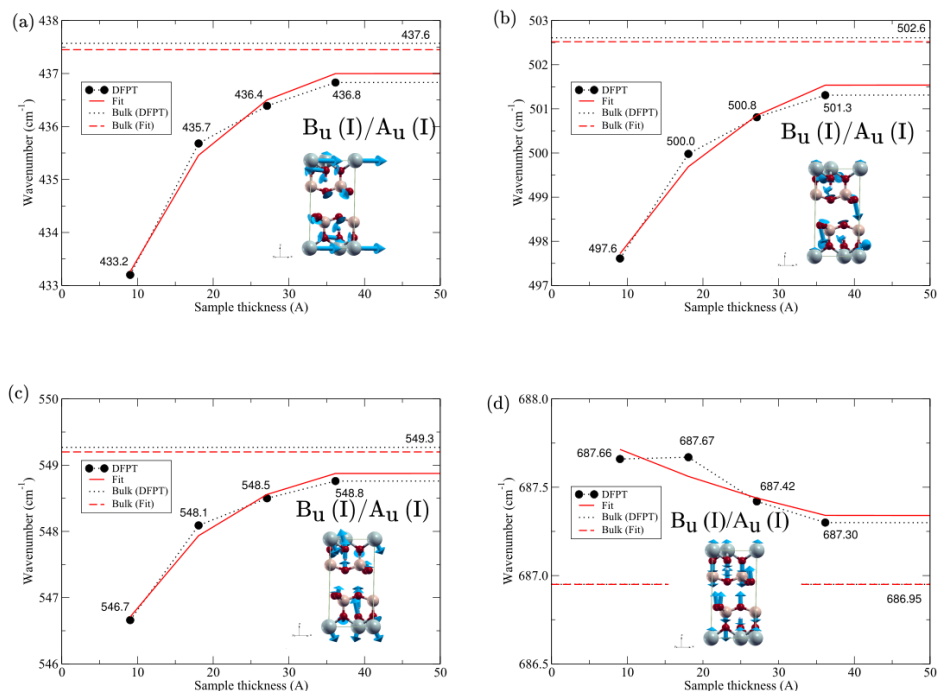

Figure S3: Wavenumber dependence with N-layer of selected infrared active modes in talc. Dots indicate wavenumbers calculated by first-principles methods. Dashed-red lines indicate the reference bulk value from first-principles calculations. Continuous-red lines indicate the fit.

In Table S3, we present the results of the semi-empirical fit, considered previously in the case of the SINS. As in the case of the SINS spectra, we found minor dependence of the wavenumber with N-layers, with is consistent with the weak interlayer interactions in talc.

*Table S3: Fitting parameters for the wavenumber dependence of selected IR-active modes of layered talc.*

| Mode [mode assignment (activity)] | $\omega_0$ (cm$^{-1}$) | $A$ (cm$^{-1}$) | $B$ (Å$^{-1}$) |
|---|---|---|---|
| 437.6 [B$_u$(I)/A$_u$(I)] | 437.45 | -8.80 | 0.082 |
| 502.6 [B$_u$(I)/A$_u$(I)] | 502.52 | -8.15 | 0.058 |
| 549.3 [B$_u$(I)/A$_u$(I)] | 549.20 | -4.90 | 0.075 |
| 686.95 [B$_u$(I)/A$_u$(I)] | 686.94 | 0.96 | 0.025 |

**Simulated Raman spectra dependence to N-layers**

Figure S3 demonstrates the influence of the dimensionality on the wavenumber of the Raman-active modes for the peaks at ~198 cm$^{-1}$, ~366 cm$^{-1}$ and ~669 cm$^{-1}$, where the N-layer wavenumbers represent the intensity-weighted average of the wavenumber of the modes observed at the vicinity of the bulk modes and we followed the atomic vibration pattern of the bulk mode down to the FL-talc regime.

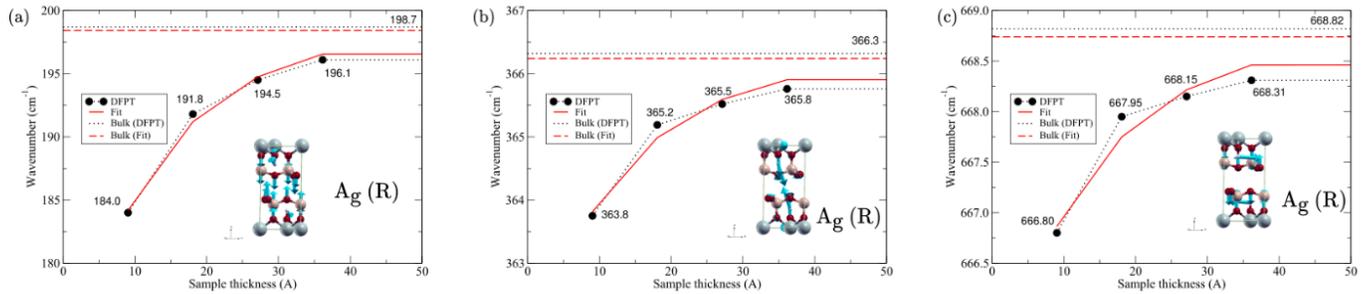

Figure S3: Wavenumber dependence with N-layer of selected Raman active modes in talc. Dots indicate wavenumbers calculated by first-principles methods. Dashed-red lines indicate the reference bulk value from first-principles calculations. Continuous-red lines indicate the fit.

In Table S4 we present the results of the semi-empirical fit, considered previously in the case of the SINS. As in the case of the SINS and IR simulated spectra, we found minor dependence of the wavenumber with N-layers, with is consistent with the weak interlayer interactions in talc.

*Table S4: Fitting parameters for the wavenumber dependence of selected Raman-active modes of layered talc.*

| Mode [mode assignment (activity)] | $w_0$ (cm$^{-1}$) | $A$ (cm$^{-1}$) | $B$ (Å$^{-1}$) |
|---|---|---|---|
| 198.7 [$A_g$ (R)] | 198.42 | -28.04 | 0.678 |
| 366.3 [$A_g$ (R)] | 366.24 | -4.69 | 0.073 |
| 668.82 [$A_g$ (R)] | 668.74 | -3.51 | 0.070 |

**References:**

[1]    Perdew J P and Wang Y 1992 Accurate and simple analytic representation of the electron-gas correlation energy *Phys Rev B* **45** 13244–9

[2]    Perdew J P, Burke K and Ernzerhof M 1996 Generalized Gradient Approximation Made Simple *Phys Rev Lett* **77** 3865–8

[3]    Perdew J P, Ruzsinszky A, Csonka G I, Vydrov O A, Scuseria G E, Constantin L A, Zhou X and Burke K 2008 Restoring the Density-Gradient Expansion for Exchange in Solids and Surfaces *Phys Rev Lett* **100** 136406

[4]    Vydrov O A and Van Voorhis T 2010 Nonlocal van der Waals density functional: The simpler the better *J Chem Phys* **133** 244103

[5]    Sabatini R, Gorni T and de Gironcoli S 2013 Nonlocal van der Waals density functional made simple and efficient *Phys Rev B* **87** 41108

[6]    Perdikatsis B and Burzlaff H 1981 Strukturverfeinerung am talk Mg3[(OH)2Si4O10] *Zeitschrift fur Kristallographie* **156** 177–86

[7]    Sabatini R, Küçükbenli E, Pham C H and de Gironcoli S 2016 Phonons in nonlocal van der Waals density functional theory *Phys Rev B* **93** 235120

[8]    Perdikatsis B and Burzlaff H 1981 Strukturverfeinerung am Talk Mg3[(OH)2Si4O,10] *Z Kristallogr Cryst Mater* **156** 177–86

[9]    Wang A, Freeman J J and Jolliff B L 2015 Understanding the Raman spectral features of phyllosilicates *Journal of Raman Spectroscopy* **46** 829–45

[10]   Rosasco G J and Blaha J J 1980 Raman Microprobe Spectra and Vibrational Mode Assignments of Talc *Appl Spectrosc* **34** 140–4

[11]   Sohier T, Gibertini M, Calandra M, Mauri F and Marzari N 2017 Breakdown of Optical Phonons' Splitting in Two-Dimensional Materials *Nano Lett* **17** 3758–63

[12]   Farmer V C 1958 The infra-red spectra of talc, saponite, and hectorite *Mineralogical Magazine and Journal of the Mineralogical Society* **31** 829–45